\documentclass[pra,aps,10pt,twocolumn,superscriptaddress,longbibliography]{revtex4-1}
\usepackage{amsfonts}
\usepackage{graphicx}
\usepackage{amsmath}
\usepackage{amssymb}
\usepackage{bm}
\usepackage{amsthm}
\usepackage{color}
\usepackage[dvipsnames]{xcolor}
\usepackage{mathrsfs,latexsym}
\usepackage{soul}
\usepackage{verbatim}
\usepackage{revsymb}
\usepackage{mhchem}
\usepackage{hyperref}
\usepackage{float}
\hypersetup{
    colorlinks=true,           linkcolor=blue,              citecolor=red,            urlcolor=blue        }

\usepackage{ulem}

\begin{document}

\title{Topological quantum interference in a pumped Su-Schrieffer-Heeger lattice}
\author{Zeng-Zhao Li}
\affiliation{Department of Chemistry, University of California, Berkeley, California 94720, USA}
\affiliation{Berkeley Center for Quantum Information and Computation, Berkeley, California 94720, USA}
\author{Juan Atalaya}
\affiliation{Department of Chemistry, University of California, Berkeley, California 94720, USA}
\author{K. Birgitta Whaley}
\affiliation{Department of Chemistry, University of California, Berkeley, California 94720, USA}
\affiliation{Berkeley Center for Quantum Information and Computation, Berkeley, California 94720, USA}

\begin{abstract}
Topological quantum interference emerges from the interplay between quantum mechanics and topology. We present evidence for two types of such interference phenomenon that can result from the quantum dynamics of initial topological states.
We realize both types of topological quantum interference in a pumped non-Hermitian Su-Schrieffer-Heeger lattice that can be implemented by creation and coherent control of excitonic states of trapped neutral atoms. 
On quenching the system from the topological to the gapless phases and then back again, 
we find that interference patterns develop in the gapless phase and also after switching back to the topological phase.
These patterns occur both as many-excitation interferences generated in the presence of pumping the atoms at the end sites, and as one- and two-excitation interferences seen in the absence of pumping when starting with edge excitations.
Investigation of the excitation dynamics shows that these interference patterns originate from the topological nature of the initial states and are very different from quantum interferences originating from non-topological states of the lattice. 
Our results also reveal that unlike well-known situations where topological states are protected against local perturbations, in the non-Hermitian SSH systems resulting from driving the excited state populations, a local dissipation at each lattice site can suppress both the topological interference and the total population of the lattice.
\end{abstract}
\date{\today}
\maketitle

\section{Introduction}

Quantum interference is a foundational concept of quantum mechanics that has given rise to a wide range of applications in quantum science and technology~\cite{ClarkeBraginski06,Scholes17nature,Stafford2007,Lambert16Physique,Samuelsson17prl,Marcos18prb,LiMartin19,Li11epl,Burkard00prb,Scholes17nature}.
In particular, quantum interference between photons is of importance for quantum communication~\cite{AzumaLo15ncomms,Yu15ncomms}, for quantum key distribution~\cite{jennewein2000quantum}, and for a range of quantum information processsing tasks, including quantum computation ~\cite{Andersen13prl} and the engineering of high-dimensional states for linear optical quantum computing~\cite{Knill01nature,Nielsen04prl,ZhangForbes16SciAdv}.  
A recent experiment~\cite{Tambasco18SciAdv} has demonstrated quantum interference between topological states of light in a photonic circuit described by the off-diagonal Harper model~\cite{Harper1955,KrausZilberberg12prl,Quandt12}.
In that work, the interference was generated by adiabatically varying the lattice coupling strength to bring topological boundary modes into spatial proximity. Here we show that it is possible to generate quantum interference of topological states in non-equilibrium physical systems under non-adiabatic conditions by making use of quantum quenches. Using this approach we address the general question of how and when topological quantum interferences can emerge from a non-trivial interplay between topology and quantum dynamics of driven systems, and examine its robustness to non-Hermitian character introduced by dissipative loss.
Using quantum quenches we can investigate interferences in both topological and non-topological phases, as well as analyze the role of different initial states in a topological phase.  In particular, given that in a topological phase there are both edge and bulk states, under a quantum quench there are several possible routes to topological quantum interference.  One is that initial edge states in a topological phase are quenched into a non-topological phase generating unconventional quantum interference.  Another is that unconventional quantum interferences develop on quenching into the topological phase.  
In this work we investigate the potential for excitonic systems to display such topological quantum interference and show that both of these routes to topological quantum interference are found under quantum quench dynamics.

A simple one-dimensional system that demonstrates topological properties is the Su-Schrieffer-Heeger (SSH) model~\cite{heeger1988solitons}. Originally proposed to explain solitonic properties of polyacetylene~\cite{SuSchriefferHeeger79prl}, this model has insulating states in the bulk but conducting edge states at the surfaces.  It has therefore been widely used in recent years as a model for topological insulators~\cite{Asboth16}.  Recent studies include experimental realizations in
waveguide photonics~\cite{Schomerus13ol,PoliSchomerus15ncomms,WeimannSzameit17nmat} and trapped atom systems~\cite{AtalaDemlerBloch13nphys,de2019observation}.
Theoretical generalizations to models with nearest-neighbour interactions~\cite{LiChen14prb,MarquesDias17prb}, to a non-Hermitian SSH model~\cite{Schomerus13ol,Lieu18prb}, and to interacting SSH chains for a Haldane phase~\cite{Sbierski18prb} have also been investigated.
SSH systems have also been studied in out-of-equilibrium situations, e.g., topological properties and phase transitions of a Floquet-engineered SSH model~\cite{GomezPlatero13prl,LagoTorres15pra,HadadAlu16prb,AtalaDemlerBloch13nphys}, Floquet-engineered topological superconductivity in a Kitaev chain~\cite{BenitoPlatero14prb,LiLamYou17prb}, and real-space effects of a quantum quench~\cite{Rossi21arXiv}.
Recently, it has been shown that pumping one end of a polaritonic SSH lattice gives rise to features of a topological phase that are dramatically distinct from those in either the trivial or gapless phases~\cite{Krivosenko18pra,Krivosenko18jpcs}. 

In this work, we propose a realization of topological quantum interference in a pumped SSH lattice. 
It has recently been shown that by tuning the phase difference of the laser fields generating the lattice potential, an optical lattice containing trapped particles can be prepared in a topologically nontrivial phase, while by further applying a pumping field to the atoms at the ends of the lattice, excitations of topological edge states can then be generated~\cite{Krivosenko18pra}.
Such excitations can be generated from diverse physical systems, such as  light-engineered polaritons in semiconductor microcavities~\cite{Amo10prb,Krivosenko18pra}, or trapped neutral atoms~\cite{de2019observation}.
We shall focus for concreteness here on the latter example of local (Frenkel) excitations of trapped neutral atoms.

In Section~\ref{sec:quench_setup} we develop protocols that coherently control both the individual atom excitations and the lattice trapping potential to generate topological quantum interferences. 
Our protocol is based on a sudden switching (``quantum quenching") of the system from the topological phase into either the gapless or the trivial non-topological phases, evolving in this phase for some time, and then switching  back. 
The first switch transforms initially localized edge state excitations of the topological phase into delocalized states responsible for the subsequent interferences, which show distinctive behavior in each of the gapless and trivial non-topological phases. 
When the system is simultaneously being resonantly driven on the atomic transitions, the excitations increase with time, producing a many-excitation interference pattern. However, when switching back to the topological phase, the localized states reappear at the ends of the lattice.

In Section~\ref{sec:quantuminterfer} we present numerical results demonstrating how this switching protocol generates the basic features of topological interference patterns. 
We find that when starting from initially localized end states, the pre-quench state in the topological phase is not an exact superposition of the two topological edge states and also possesses small components from bulk states, giving non-zero amplitude on the bulk sites. 
This topological superposition state evolves under the topological phase Hamiltonian until the first switch, which is less noticeable in the pumped case than in the non-pumped case since the pumping fields needs time to build up the edge-site occupations while the initial occupation in the case of no pump is ready to decrease into the bulk.
After the first switch into the gapless phase, 
we find that the delocalization of edge excitations in the topological phase provides the main contribution to the interference in the gapless phase. 
When switching back to the topological phase, in addition to repopulating the topological edge states with higher values of occupation number on account of the applied pumping fields, we observe a second interference pattern that is distinct from that seen in the gapless phase.

In order to distinguish topological interference patterns from  non-topological patterns, we compare the quantum interference excitation patterns with the excitation patterns obtained for evolution in the topological nontrivial and gapless phases under pumping but in the absence of any lattice potential phase switching. 
In the topological regime excitations are always localized at the edges and cannot interfere with each other, but in the gapless regime interferences can develop under pumping due to the formation of delocalized states.  Here under pumping of both end sites we find interference patterns that reveal an interplay between the patterns determined by 
pumping the even and odd sublattices separately. 
We also consider the switching protocol under a single pumping field, which shows the expected interference pattern derived from pumping a single sublattice.

In Section~\ref{sec:double_nopump} we explore the time evolution of the lattice excitation with the pumping fields switched off and following instead two initial excitations, one at the either end of the lattice, 
while in Section~\ref{sec:total+dissipation} we investigate the time evolution of the entire lattice population and investigate the effect of local dissipation at individual lattice sites.   Here, contrasting with the known robustness of topological states against local perturbations in a non-driven situation, we find that both the total population of excitations in the topological phase and the observed topological quantum interference can be suppressed by the presence of local dissipation at each lattice site. 

This analysis of quantum interference in a driven quantum system possessing topological states under quantum quench scheme reveals remarkable features distinct from those in purely topological, gapless or trivial non-topological phases. 
Findings such as the delocalization of exponentially localized topological edge states into the bulk responsible for the observed unexpected interference after implementing a quantum quench, can enhance our understanding of emergent phenomena resulting from a non-trivial interplay among topology, non-Hermitian character, and quantum quenches in non-equilibrium systems. 
This work explores new applications of topology in the dynamics of interference-based quantum technology and devices.

\begin{figure}
\centering
  \includegraphics[width=.99\columnwidth]{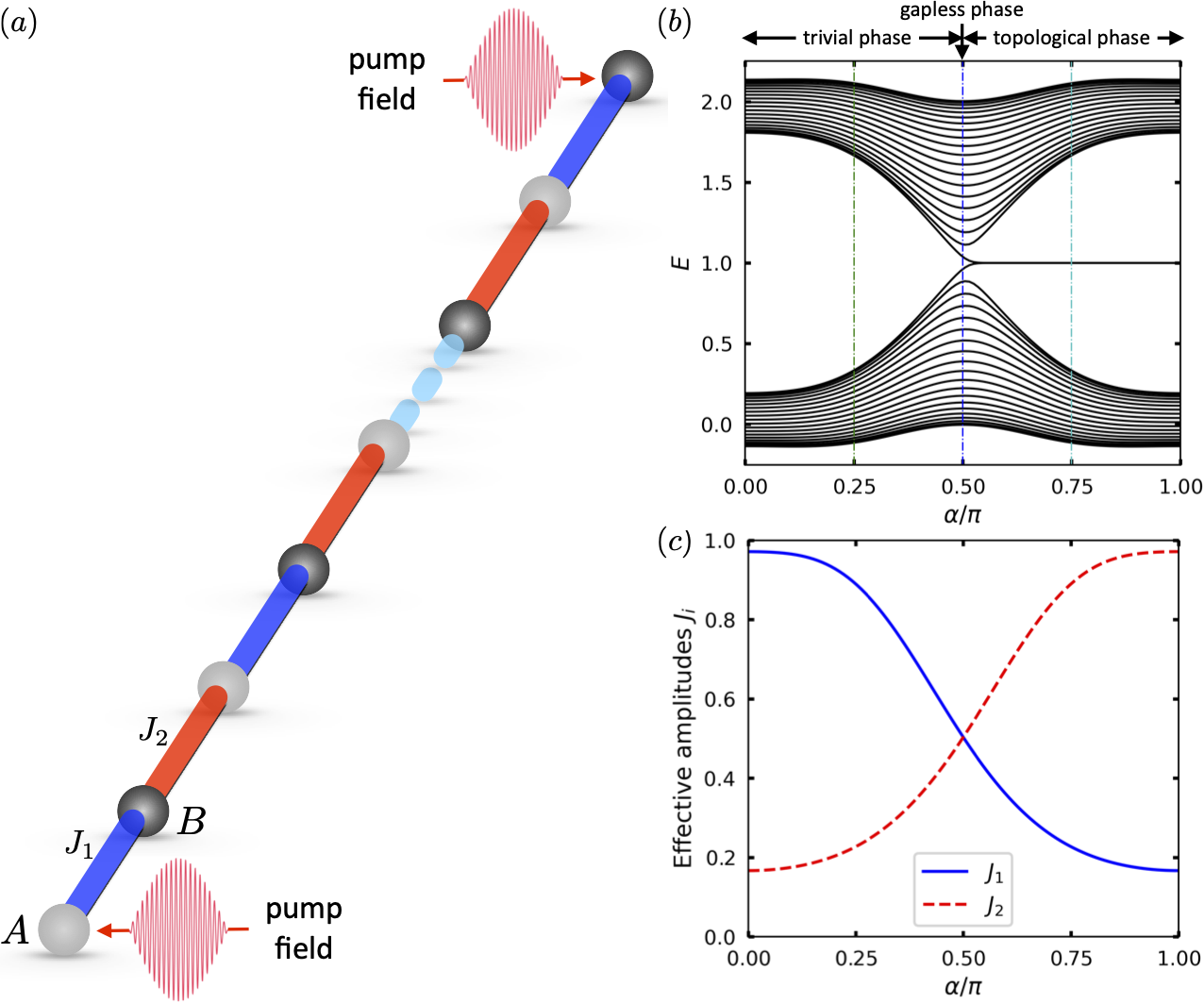}
\caption{(color online) 
(a) Schematic diagram of the SSH lattice with pumping fields applied at both end sites.
(b) Energy spectrum plotted as a function of the parameter $\alpha$ specifying the relative phase of the three laser fields generating the lattice potential. Vertical green, blue, and cyan lines denote the topologically trivial ($\alpha=0.25\pi$), gapless ($\alpha=0.5\pi$), and topological ($\alpha=0.75\pi$) phases, respectively. 
(c) Effective intra- and inter-cell  hopping amplitudes $J_1$ and $J_2$ as a function of $\alpha$. 
The other parameter values for evaluating energy spectrum of $H_{\rm SSH}$ in Eq.~(\ref{eq:hamiltonian}) and hopping amplitude in Eq.~(\ref{eq:hopping}) are $\varepsilon=1$, $\gamma=0$, $N=20$, $V_0=0.125$, and $\mu/k^2=0.25$. }
\label{fig:schematicSpectrAmplit}
\end{figure}

\section{Sudden Switching of a pumped non-Hermitian SSH lattice}
\label{sec:quench_setup}
\subsection{Model for pumped non-Hermitian SSH lattice}
\label{sec:tunephase}

The pumped non-Hermitian SSH lattice that we study here is schematically shown in Fig.~\ref{fig:schematicSpectrAmplit}(a).
In the absence of pumping fields, the finite non-Hermitian SSH system is similar to the original SSH model~\cite{SuSchriefferHeeger79prl}, extended here to include an onsite energy $\varepsilon$, representing, for example, {specific site energy levels in a trapped neutral atom implementation or} a cavity energy in photonic implementations~\cite{Han19}
and a local dissipation term characterized by $\gamma$ that accounts for the decay of excitations at each lattice site. 
For trapped neutral atoms this could represent spontaneous emission, or vibrational dissipation for atoms trapped in an excited state of the lattice potential.

The Hamiltonian describing this static system is
\begin{eqnarray}
H_{\rm SSH} &=& \sum_{l=1}^N J_1(|l,A\rangle\langle l,B| +|l,B\rangle\langle l,A|) \notag\\
&&+\sum_{l=1}^N J_2 (|l,B\rangle\langle l+1,A| +|l+1,A\rangle\langle l,B|) \notag\\
&&+ (\varepsilon - i\gamma) \sum_{l=1}^N (|l,A\rangle\langle l,A| + |l,B\rangle\langle l,B|),
\label{eq:hamiltonian}
\end{eqnarray}
where $N$ is the number of unit cells, each consisting of $A$ and $B$ sites, represented by grey and black dots respectively in Fig.~\ref{fig:schematicSpectrAmplit}(a). The coefficients $J_1$ and $J_2$ are intra- and inter-cell hopping amplitudes, represented by blue and red bonds respectively.
For $J_1\gg J_2$, the $N$ unit cells are only weakly coupled to each other and excitations are almost localized in individual unit cells.
This corresponds to the trivial non-topological phase which is characterized by a finite gap. 
For $J_1\ll J_2$, the unit cells are effectively reorganized, with $J_1$ and $J_2$ now becoming the inter- and intra-cell hopping amplitudes respectively.
The last site at each end of the lattice is then an edge site in an incomplete unit cell.
This is the topological phase in which there exist two degenerate energy states localized on these edge sites that can be entangled across the chain. These edge states are energetically located within a band gap separating delocalized states.  The topological phase is thus characterized by the presence of the localized edge states within a finite band gap.
We shall also explicitly distinguish the unique gapless phase, which for the infinite system is located at the special point $J_1 = J_2$, and where the gap is equal to zero but there are no localized edge states.
In both the trivial non-topological phase and the gapless phase, the eigenstates are all typical Bloch states, in which an intra-unit cell Bloch function consisting of symmetric or antisymmetric superposition of the two site localized states is modulated by a plane wave of fixed $k$.  The additional localized edge states that are formed only in the topological phase consist of Bell state pairs formed by the two edge states on opposite ends of the lattice and vanishing amplitude in the interior~\cite{Suppl}.

The topological phase transition from the trivial phase to the topological phase can be induced by tuning the hopping amplitudes, either going continuously through the gapless phase or by a discontinuous sudden change or ``quench" of these, as discussed in detail below.  
We focus here on the consequences of the transition for a pumped non-Hermitian SSH lattice.
In particular, we investigate the effects of non-equilibrium driving of the system from the ground state in the topological phase, corresponding to the half-filled SSH chain of $2N$ sites with $N$ electrons populating $N-1$ bulk dimeric states in the interior of the chain and $1$ electron in one of the two entangled edge states.  The non-equilibrium driving will be achieved by pumping one or both of the two edge sites.

{To realize $H_{\rm SSH}$ in a trapped neutral atom system, we take advantage of the possibility to control the hopping amplitudes $J_1$ and $J_2$ in such systems}~\cite{AtalaDemlerBloch13nphys}. Considering an optical potential $V_{OL}(x,\tau) = V_0 [3+4\cos(2kx)\cos(\alpha(\tau))+2\cos(4kx)]$ generated by three laser fields with amplitudes proportional to $e^{i(kx+\alpha)}$, $e^{-ikx}$, and $e^{i3kx}$, respectively, and applying the harmonic approximation around the local minima of $V_{OL}(x,\tau)$, the hopping amplitudes can be expressed as~\cite{Krivosenko18pra,Suppl}
\begin{eqnarray}
J_i &=& \frac{\omega}{2} e^{-\Delta_i^2} (\Delta_i^2 + \frac{1}{2}),
\label{eq:hopping}
\end{eqnarray}
where
\begin{eqnarray}
\omega &=& \sqrt{\frac{8V_0k^2}{\mu} (4-\cos^2\alpha)} , \\
\Delta_1 (\alpha) &=& \arccos\frac{\cos\alpha}{2} [\frac{8V_0\mu}{k^2}(4-\cos^2\alpha)]^{1/4}, \label{eq:Delta1} \\
\Delta_2 (\alpha) &=& \Delta_1 (\pi-\alpha), \label{eq:Delta2}
\end{eqnarray}
with $\omega$ and $\mu$ the vibrational frequency and reduced oscillator mass, respectively.

\begin{figure}
\centering
  \includegraphics[width=.99\columnwidth]{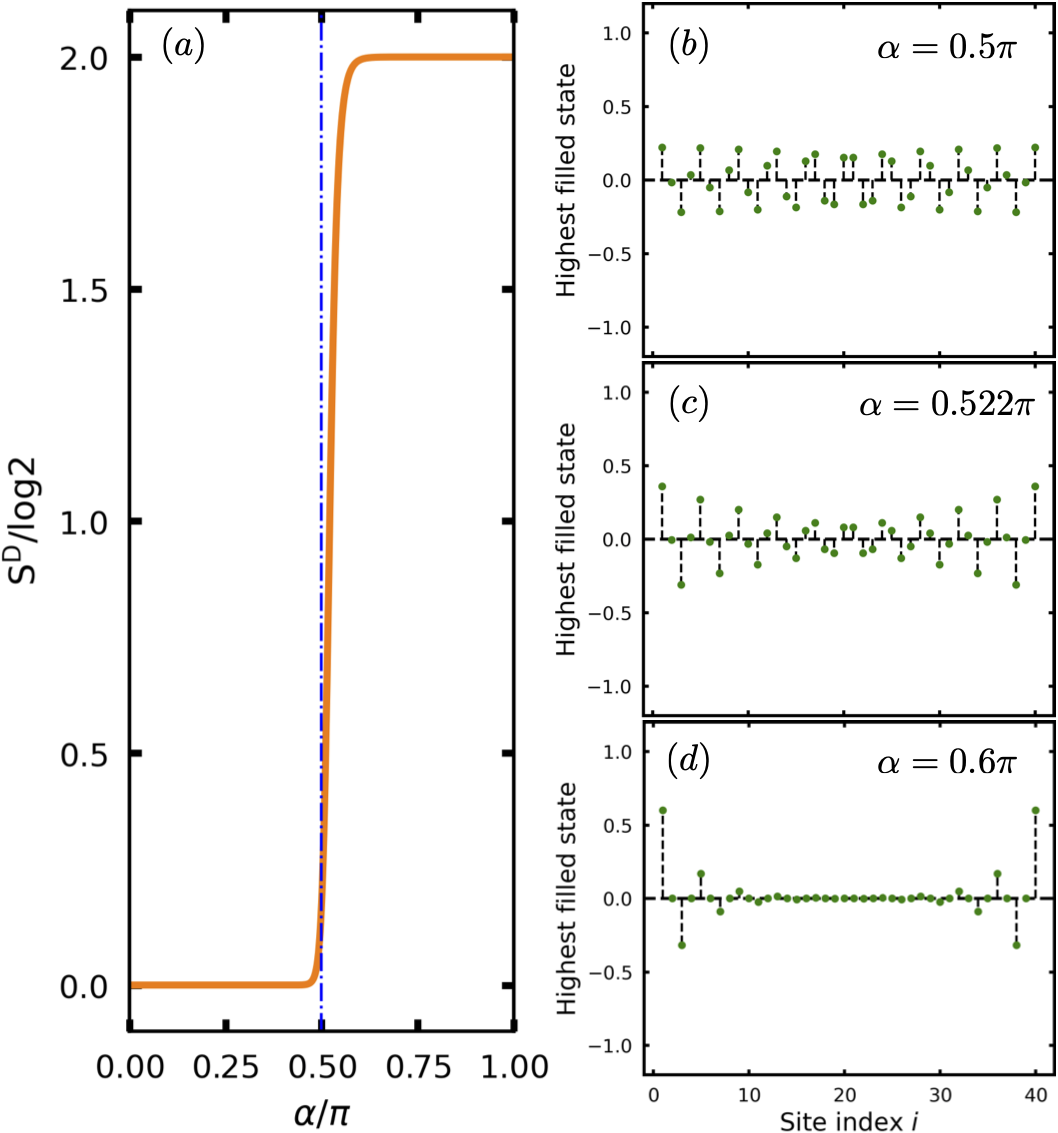}
\caption{(color online)
{(a) Disconnected entanglement entropy $S^D$ as a function of $\alpha$, for the ground state of the SSH model at half filling on a chain of $2N=40$ sites. $S^D$ is evaluated by dividing the $2N$ sites evenly into four non-overlapping segments $A, B, D$, and $C$, each having $N/2$ sites, then evaluating $S^{D}=S_{AB}+S_{BC}-S_{ABC}-S_{B}$, where $S_i=-{\rm Tr}_i\rho_i \log(\rho_i)$ the von Neumann bipartite entanglement entropy.}
The panels on the right show the distribution of the amplitudes of the highest filled eigenstates over all sites for (b) $\alpha=0.4\pi$, (c) $\alpha=0.5\pi$, and (d) $\alpha=0.522\pi$. 
All parameters are the same here as in Fig.~\ref{fig:schematicSpectrAmplit}.}
\label{fig:SD}
\end{figure}

The relative phase $\alpha$ of the laser field components
is then used to tune the properties of the lattice as in~\cite{AtalaDemlerBloch13nphys}. 
We consider $2N=40$ sites as an example, employing open boundary conditions.
The  hopping amplitudes and the energy band structure are shown as a function of the relative phase $\alpha$ in Fig.~\ref{fig:schematicSpectrAmplit}(c) and (b), respectively.
Fig.~\ref{fig:schematicSpectrAmplit}(c) shows that in the regime $0\le\alpha<0.5\pi$, $J_1$ is larger than $J_2$.  Fig.~\ref{fig:schematicSpectrAmplit}(b) shows that this regime is characterized by {an energy gap with no localized states in the gap, indicating the topologically trivial quantum phase.}
The gap decreases as $\alpha$ approaches $\pi/2$. At this point the hopping amplitudes are equal, $J_1=J_2$, and {in the finite system shown in Fig.~\ref{fig:schematicSpectrAmplit}(a) the energy gap is almost zero.  In the limit of an infinite chain the gap is identically zero and the system is strictly gapless.}
On further increase of $\alpha$, i.e., $\alpha>0.5\pi$, the hopping amplitudes are reversed in order, i.e., now $J_1<J_2$. 
Fig.~\ref{fig:schematicSpectrAmplit}(b) shows that two degenerate energy states now appear in the gap (with finite energy because of the nonzero value of the onsite energy $\varepsilon=1$ in this example.) These correspond to  edge states that are decoupled from the rest of the lattice.  The system is now in the topologically non-trivial phase.
Thus Fig.~\ref{fig:schematicSpectrAmplit}(b) and (c) shows that the SSH lattice can be tuned between  topologically trivial and nontrivial phases by changing the relative phase $\alpha$ of the lasers whose interaction with the neutral atoms defines the optical lattice~\cite{AtalaDemlerBloch13nphys}.

To characterize the finite size analog of the bulk topological phase transition we employ here the disconnected entanglement entropy $S^D$ introduced in~\cite{ZengWen19,Micallo20spp}.  $S^D$ is defined in terms of both connected and disconnected entanglement entropies of different segments of the system, in order to cancel out all area and volume law contributions to the bipartite entanglement entropy.  It is thus sensitive only to the non-local topological entanglement stored within the ground state manifold and acts as an effective order parameter. 
Fig.~\ref{fig:SD}(a) shows the $\alpha$ dependence of $S^D$. For the finite SSH chain in the topological phase, with both periodic and open boundary conditions at large chain length, $S^D$ is equal to $2 \log 2$ as a result of the entanglement of the localized edge states across the full extent of the chain. 
With open boundary conditions, $S^D$ is equal to zero only in the trivial gapped phase, with a smooth transition between the two phases that becomes increasingly abrupt as the chain size increases~\cite{Micallo20spp}.  Evaluating $S^D$  using the free fermion techniques of ~\cite{Peschel03jpa} as outlined in Ref.~\onlinecite{Micallo20spp}, yields the behavior in Fig.~\ref{fig:SD}(a) for our chain with $2N=40$ sites. This shows a clear separation of the two phases, with the onset of a sharp rise starting just before the bulk critical point at $\alpha=\pi/2$.  For this particular finite size chain we see that $S^D/\log 2 = 1$ at $\alpha \simeq 0.522\pi$.  However, 
inspection of the wavefunctions for the values $\alpha=\pi/2$ and $\alpha =0.522 \pi$ (right panels in Fig.~\ref{fig:SD}) show that the highest filled state is approximately uniformly delocalized over the full chain for $\alpha =\pi/2$ while it starts to develop maximal amplitudes at the edges already for $\alpha =0.522 \pi$. We therefore take the point $\alpha = \pi/2$ as representative of the gapless phase in this finite sized system and interpret the shift of the $S^D=1$ location to $\alpha=0.522\pi$ as a finite size effect.

We consider here the effects of coherently pumping one or both of the edge sites of the ground state of the SSH system in the topological phase, i.e., for $\alpha > 0.5\pi$.  We employ a semiclassical treatment~\cite{Pervishko13oe}, applying site-specific laser driving terms $F_{A}e^{-i(\omega_{pA} t-\phi_{0A})}$ and $F_{B}e^{-i(\omega_{pB} t-\phi_{0B})}$ at the first and last sites of the lattice, respectively, as illustrated schematically in Fig.~\ref{fig:schematicSpectrAmplit}.
The terms $F_{A(B)}$, $\omega_{pA(B)}$, and $\phi_{0A(B)}$ denote respectively the amplitude, frequency, and relative phase of the pumping field at the first(last) site.
The dynamics of the pumped SSH lattice can then be described by the following equation (see Methods for detailed analysis)
\begin{eqnarray}
i\dot{x}&=&H_{\rm SSH} x + f,
\label{eq:equation}
\end{eqnarray}
where $x$ represents the amplitude vector of the lattice excitations and $f$ the pumping field. The excitation amplitude vector $x$ is given by $x=(A_1,B_1,\cdots,A_N,B_N)^T$, with $A_l$ and $B_l$ denoting the time-dependent amplitudes of the states $|l,A\rangle$ and $|l,B\rangle$ respectively, [see Eq.~(\ref{eq:hamiltonian})].
For pumping at the two ends of the lattice, the pumping amplitude vector $f$ is given by $f=(F_A e^{i(\phi_{0A}-\omega_{pA} t)},0,\cdots,0,F_Be^{i(\phi_{0B}-\omega_{pB} t)})^T$.
Specialization to a single pumping field at the first or last site corresponds to taking $F_B=0$ or $F_A=0$, respectively. 
This could be realized, e.g., by placing the optical lattice inside a microcavity~\cite{Amo10prb,Pervishko13oe,Krivosenko18pra}.  In this work we also include a dissipation term $-i\gamma$ at each site, which allows the effects of e.g., spontaneous emission to be taken into account for trapped neutral atoms. 

To demonstrate the formation of an interference pattern, we consider the excitation intensity to be represented by the occupation number  $P_{\rm pop}$ on each lattice site,  where $P_{\rm pop}=|A_l(t)|^2(|B_l(t)|^2)$ for $A(B)$ sites, respectively. Then the total excitation intensity or population of excitations over the whole lattice is given by $P_{\rm tot}=\sum_l (|A_l(t)|^2+|B_l(t)|^2)$.
Resonant pumping is achieved when the pumping frequency satisfies $\omega_p=\omega_{pA}=\omega_{pB}=\varepsilon$.
In addition, we impose the symmetry contraint $\phi_{0A}=\phi_{0B}=\phi_{0}$ on the phases of the pumping fields, which enforces indistinguishability of the quantum states (i.e., $|1,A\rangle$ and $|2N,B\rangle$) and will enhance the appearance of any quantum interference.

We note that the SSH Hamiltonian, Eq.~(\ref{eq:hamiltonian}) is a single-particle Hamiltonian and is thus directly applicable to a situation of the non-pumped system with single excitations. When considering a pumped SSH lattice with multiple excitations, the effects of interactions between excitations may become important.  Here we regard these as being included within a mean-field analysis in which any interactions between excitations are already included in the parameters of Eq.~(\ref{eq:hamiltonian}).  This assumes that 
the particles are independent of each other (consistent with bosonic excitations) and that the effect of all other excitations on any given excitation is approximated by averaging over the interactions to modify, e.g., the on-site energy in Eq.~(\ref{eq:hamiltonian}). 

\subsection{Interference of topological states induced by phase switching}
\label{sec:switch}

It was recently shown that topological states can undergo quantum interference~\cite{Tambasco18SciAdv}. The first experimental demonstration of interfering topological states of light was performed in a photonic waveguide circuit, which can be modeled by the off-diagonal Harper model~\cite{Harper1955,KrausZilberberg12prl}.
We now show how to achieve such topological interference by quenching a pumped SSH lattice.

Our protocol relies on sudden switching between two phases, namely, between the topologically nontrivial and gapless phases, by changing the relative phase difference $\alpha$ of the laser fields that generate the lattice potential.
Specifically, we first prepare two topological edge states by tuning the SSH lattice into the topological phase via setting $\alpha$ to $\alpha_{\rm T}=0.75\pi$  [see Fig.~\ref{fig:schematicSpectrAmplit}(b)].
These edge states are confined to the edges in the topological phase. To delocalize them, we then switch to the gapless phase at time $t_a$ by resetting $\alpha$ to $\alpha_{\rm G}=0.5\pi$. 
With this quantum quench, i.e., a sudden change of the Hamiltonian to the one that hosts gapless states, one would expect that the topological states then become  mobile and interfere with each other.
Alternatively, one can also switch to the trivial phase via setting $\alpha$ to $\alpha_{\rm tr}=0.25\pi$ to realize the delocalization of the topological states. We also consider this case below.

Under this scheme, the Hamiltonian in Eq.~(\ref{eq:equation}) becomes 
$H_{\rm SSH} (\alpha(t))$ with
\begin{eqnarray}
\alpha(t) &=& \Biggr\{ \begin{array}{lll}
\alpha_{\rm T} & \text{ if } 0\le t\le t_a \\
\alpha_{\rm G} & \text{ if } t_a< t\le t_b \\
\alpha_{\rm T} & \text{ if } t_b< t
\end{array}
\label{eq:scheme}
\end{eqnarray}
where $\alpha_{\rm T}=0.75\pi$ and $\alpha_{\rm G}=0.5\pi$ [see the cyan and blue vertical lines in Fig.~\ref{fig:schematicSpectrAmplit}(b)]. In the calculations below we take $t_a=10T_p$, and $t_b=30T_p$, where $T_p=2\pi/\omega_p$, with $\omega_p$ the frequency of the pumping fields.
We mention that all energy quantities and time variable are defined in units of $\omega_p$ and $T_p$, respectively. 
For the case of the non-pumped system that doesn't involve frequency or period of pumping fields, we alternatively introduce a new variable $J(=\omega_p)$ as the energy unit (for e.g., $\varepsilon, J_1, J_2, \gamma$) to enable the protocol in Eq.~(\ref{eq:scheme}) to work as well and meanwhile to avoid anything misleading. That leads to $Jt_a/2\pi=10$ and $Jt_b/2\pi=30$ for the case of no pump.

\begin{figure}
\centering
  \includegraphics[width=.96\columnwidth]{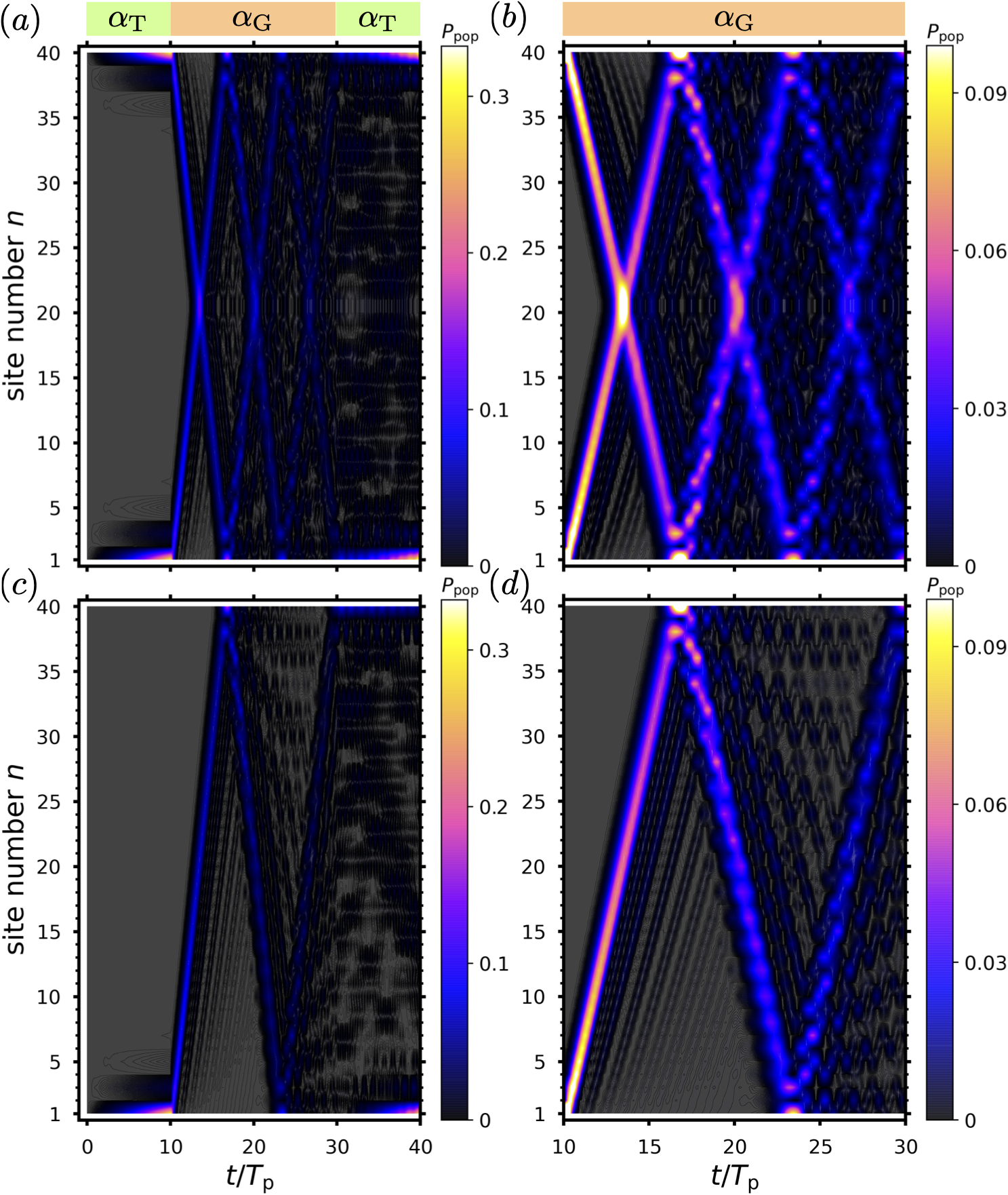}
\caption{Time-dependent occupation number $P_{\rm pop}$ of the SSH lattice during an interference inducing quantum quench protocol under pumping at both ends ($F_{\rm A}=F_{\rm B}=0.01$, upper panels), and under pumping at one end only ($F_{\rm A}=0.01$, $F_{\rm B}=0$, lower panels). The initial lattice condition at $t=0$ is no excitation, i.e., $A_l(0)=B_l(0)=0$. 
Topological interference is observed in panels (a) and (c) with parameters $\alpha_{\rm T}=0.75\pi$, $\alpha_{\rm G}=0.5\pi$, and switching time interval specified by $t_a=10T_p$, $t_b=30T_p$.
Panels (b) and (d) are zoom-in views of the patterns in the gapless phase quench period $10T_p<t<30T_p$ of panels (a) and (c), respectively.
The other parameters used are $\gamma=0.0025$, $N=20$, $\varepsilon=1$, $V_0=0.125$, $\mu/k^2=0.25$, $\omega_{pA}=\omega_{pB}=\omega_{p}=1$, $\phi_{0A}=\phi_{0B}=\phi_0=0$. 
All energy quantities are defined in units of $\omega_p$ if not specified.
}
\label{fig:multiparticle}
\end{figure}

\section{Many-excitation quantum interference induced by pumping end sites of the SSH chain}
\label{sec:quantuminterfer}

Fig~\ref{fig:multiparticle} shows the time-dependence of the site occupations under the phase switching scheme described above, in the presence of pumping either both end sites (top panels) or one end site only (bottom panels). %
The generation of interference patterns when switched into the gapless phase from the topological phase is evident for both forms of pumping.

Panel (a) of Fig.~\ref{fig:multiparticle} shows the behavior under pumping at both ends.  The system is initially located in the topological phase with $\alpha_{\rm T}=0.75\pi$ and remains there for $t\le t_a=10T_p$. Here the largest population (indicated in purple) is strongly localized at the edges, i.e., at the first and the last sites, signifying the successful generation of topological edge states.
In the immediate aftermath of time $t_a$ the system parameters are suddenly switched to $\alpha_{\rm G}=0.5\pi$ and one would have expected the topological edge states start to delocalize, after which they rapidly get close enough to interfere with each other in the bulk region of the lattice.
Immediately after $t_b=30T_p$ the system is switched back to the topological phase by setting $\alpha_{\rm T}=0.75\pi$ and after this time topological states reappear at ends of the lattice. These states are much more highly excited than before. This is due to the additional excitations that were added by the pumping fields at the two end sites of the lattice. These excitations persist for long durations.
To more clearly reveal the interference pattern,  Fig.~\ref{fig:multiparticle}(b) shows a zoom-in view of  Fig.~\ref{fig:multiparticle}(a) for the period $10T_p<t\le30T_p$ during which the system is in the gapless regime. The interference pattern is seen to be symmetric with respect to the center of the lattice.
The sharpness of the pattern decreases with time, due to the damping effects (here $\gamma=0.0025$) that dissipate the excitations at each site.

Panel (c) of Fig.~\ref{fig:multiparticle} shows the behavior under pumping of only the first site. In this case a large occupation is achieved only at the first site.  
The corresponding zoom-in view of the patterns in the period $10T_p<t\le 30T_p$ is presented in Fig.~\ref{fig:multiparticle}(d) where the lack of symmetry with respect to the center of the lattice is now evident.
The case of pumping the last site is considered as well~\cite{Suppl}.

\begin{figure}
\centering
  \includegraphics[width=1.0\columnwidth]{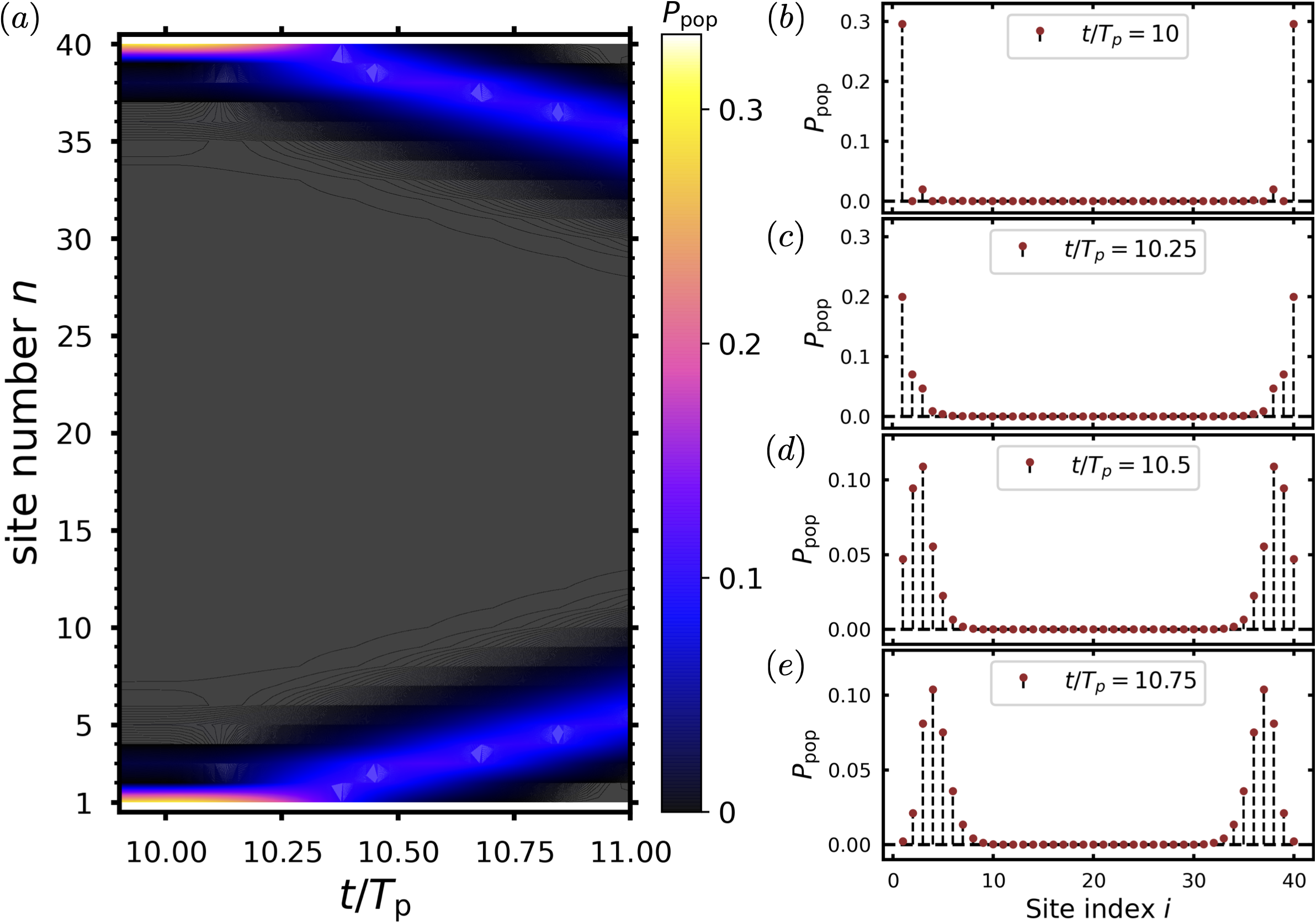}
\caption{
(a) A zoom-in view of Fig.~\ref{fig:multiparticle}(a) around the first switch at $t/T_p=10$. The occupation number $P_{\rm pop}$ across the SSH lattice when (b) $t/T_p=10$, (c) $t/T_p=10.25$, (d) $t/T_p=10.5$, and (e) $t/T_p=10.75$.
}
\label{fig:multiparticle_quenchHysteresis}
\end{figure}

An interesting feature of  Fig.~\ref{fig:multiparticle} is that although the system is switched into the gapless phase at $t=t_a=10T_P$, the delocalization of the edge states appears to start from the later times, for both the single and double end pumping. 
This hysteresis, as a phenomenon in which the post-quench state lags behind the change of the relative phase $\alpha$, is more obvious in Fig.~\ref{fig:multiparticle_quenchHysteresis}(a) (i.e., a zoom-in view of Fig.~\ref{fig:multiparticle}(a) around the first switch).
A further demonstration is seen from comparing the lattice occupation number at e.g., $t=10T_p$ (immediately before the quench), and $t=10.25T_p, 10.5T_p, 10.75 T_p$ (after the quench) as shown in Fig.~\ref{fig:multiparticle_quenchHysteresis}(b) and (c), (d), (e), respectively. 
This observation derives basically from the fact that $J_1/\omega_p=0.5$ in the gapless phase which leads to $1/J_1\sim 0.318T_p$ as a time scale for the population at the edges to delocalize into the bulk.

\begin{figure}
\centering
  \includegraphics[width=.98\columnwidth]{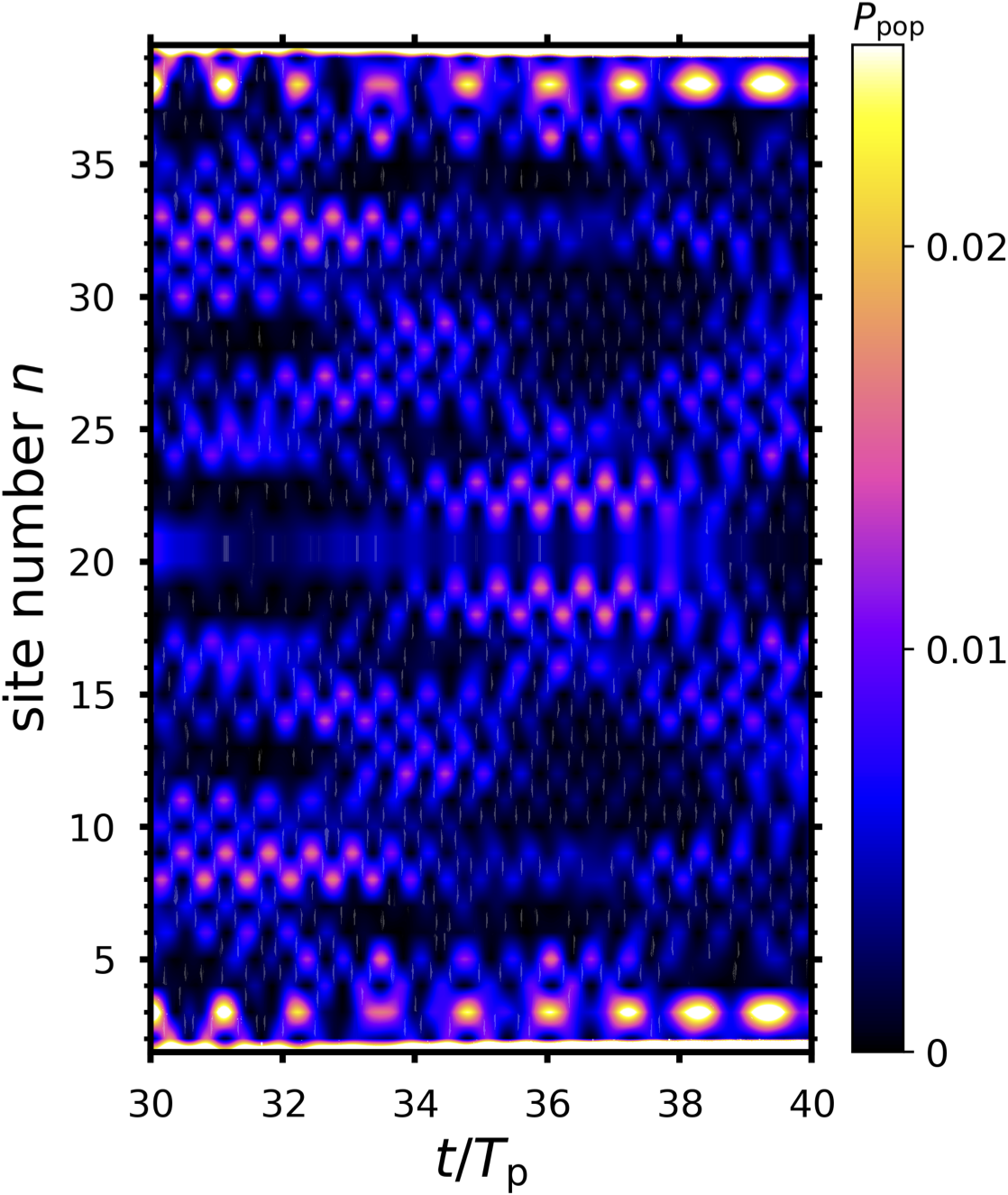}
\caption{A zoom-in view of the interference pattern in the topological phase quench period $30T_p<t\le40T_p$ of Fig.~\ref{fig:multiparticle}(a). 
The plot shows the occupation number $P_{\rm pop}$ of the SSH lattice with quantum quench protocol under pumping at both ends ($F_{\rm A}=F_{\rm B}=0.01$, upper panels), with initial lattice condition at $t=0$ given by $A_l(0)=B_l(0)=0$ (no excitation). 
Other parameters are the same as in Fig.~\ref{fig:multiparticle}(a), i.e., $\alpha_{\rm T}=0.75\pi$, $\alpha_{\rm G}=0.5\pi$, $t_a=10T_p$, $t_b=30T_p$, 
$\gamma=0.0025$, $N=20$, $\varepsilon=1$, $V_0=0.125$, $\mu/k^2=0.25$, $\omega_{pA}=\omega_{pB}=\omega_p=1$, $\phi_{0A}=\phi_{0B}=\phi_0=0$.}
\label{fig:multiparticle_A_plus}
\end{figure}

We also see an interference pattern after switching back to the topological phase in Fig.~\ref{fig:multiparticle}(a).
Its zoom-in view is presented in Fig.~\ref{fig:multiparticle_A_plus}, from which it is clear that this is a different pattern than that of Fig.~\ref{fig:multiparticle}(b).
As expected given that the pumped fields are applied at the end sites, we see that the maximum occupation is at these end sites (see Fig.~\ref{fig:multiparticle}(a)). 
The occupation probability then  decreases for the sites away from the edges and an interference pattern appears in the bulk which is quite different from the pattern obtained after switching from the topological to the gapless phase.

\begin{figure}
\centering
  \includegraphics[width=.96\columnwidth]{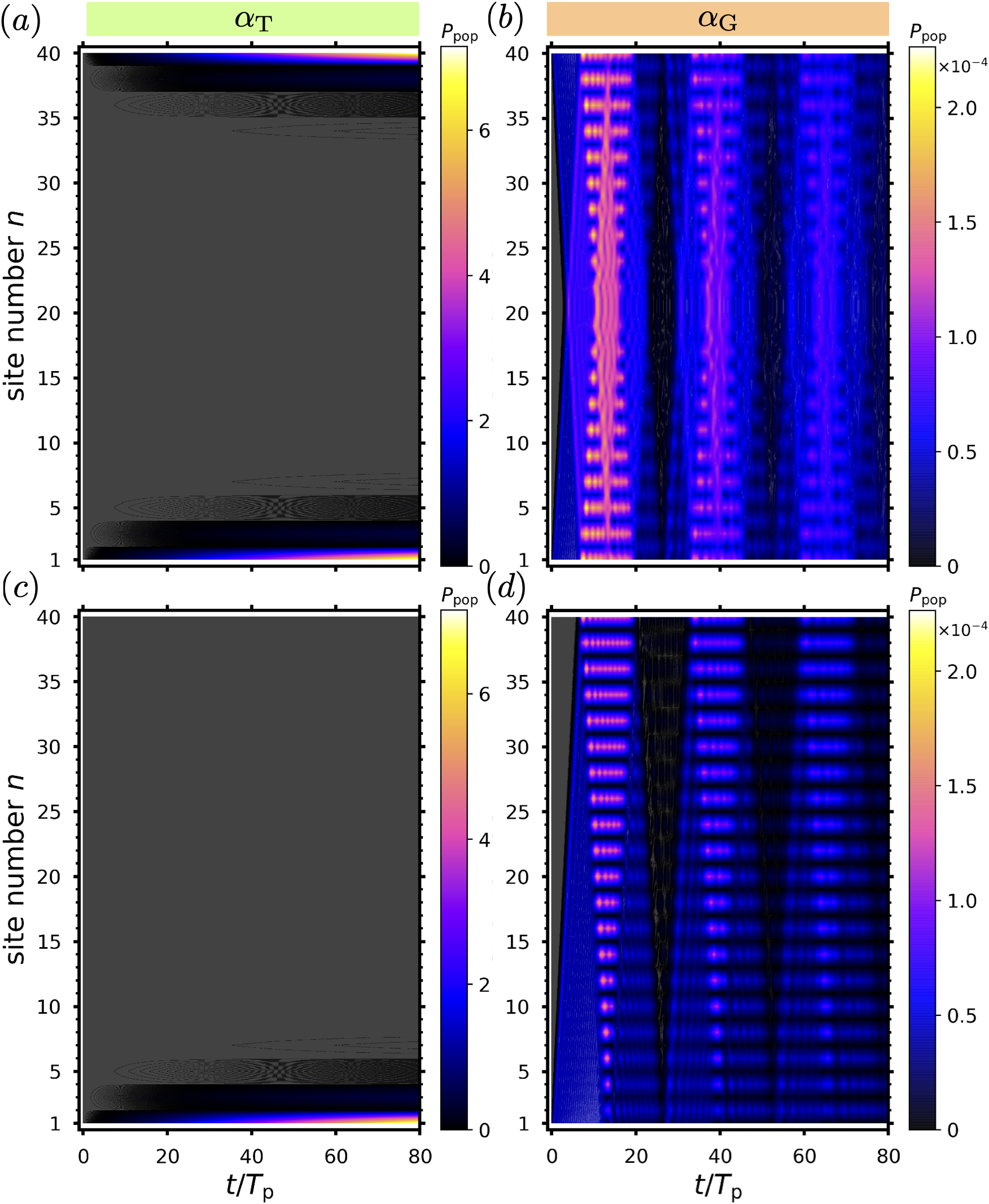}
\caption{
Time-dependent occupation number $P_{\rm pop}$ of the SSH lattice in the topological phase ($\alpha_{\rm T}=0.75\pi$, left panel) and in the gapless phase ($\alpha_{\rm G}=0.5\pi$, right panel) under pumping at both ends ($F_{\rm A}=F_{\rm B}=0.01$, upper panels) and under pumping at one end only ($F_{\rm A}=0.01$, $F_{\rm B}=0$, lower panels). No quench is applied here.
The initial condition here is no excitation at time $t=0$, i.e., $A_l(0)=B_l(0)=0$ (same initial condition as in Fig.~\ref{fig:multiparticle}). All other parameters are the same as in Fig.~\ref{fig:multiparticle}.}
\label{fig:multiparticle_B}
\end{figure}

To distinguish between these interference patterns that have their origins in topology from interferences derived from conventional non-topological states,
we now consider the topologically nontrivial and the gapless phases separately and omit phase switchings between them. Fig.~\ref{fig:multiparticle_B} summarizes the dynamical results for the topologically nontrivial phase in the left panels and for the gapless phase in the right panels.

In the topologically nontrivial phase, on pumping at both ends of the lattice one observes a rapid increase of the occupation number at each end of the lattice, as seen in Fig.~\ref{fig:multiparticle_B}(a). It is obvious that under these conditions, no interference can occur between localized states that are far away from each other.
The absence of an interference in Fig.~\ref{fig:multiparticle_B}(a) also implies that the two edge states are independent of each other and can be obtained by separately pumping either the first [see Fig.~\ref{fig:multiparticle_B}(c)] or the last site [see Fig.~S3(c) in Ref.~\onlinecite{Suppl}]. Thus one cannot obtain the interference of topological states merely by driving the system in the topological phase.

In the gapless phase, the excitations behave quite differently. 
Let us start from the case of a single pumping field at the first site, shown in Fig.~\ref{fig:multiparticle_B}(d). High occupations at even sites away from the pumped site are observed.
Conversely, using instead the second pumping field at the last site gives rise to high occupations at odd sites, i.e., the mirror image under reflection symmetry with respect to the center of the lattice [see Fig.~S3(d) in Ref.~\onlinecite{Suppl}]. It is then evident that the symmetric pattern seen for pumping a both ends in Fig.~\ref{fig:multiparticle_B}(b) is generated by an interference of the amplitude derived from pumping at the first site with that derived from pumping at the last site.

\begin{figure}
\centering
  \includegraphics[width=.98\columnwidth]{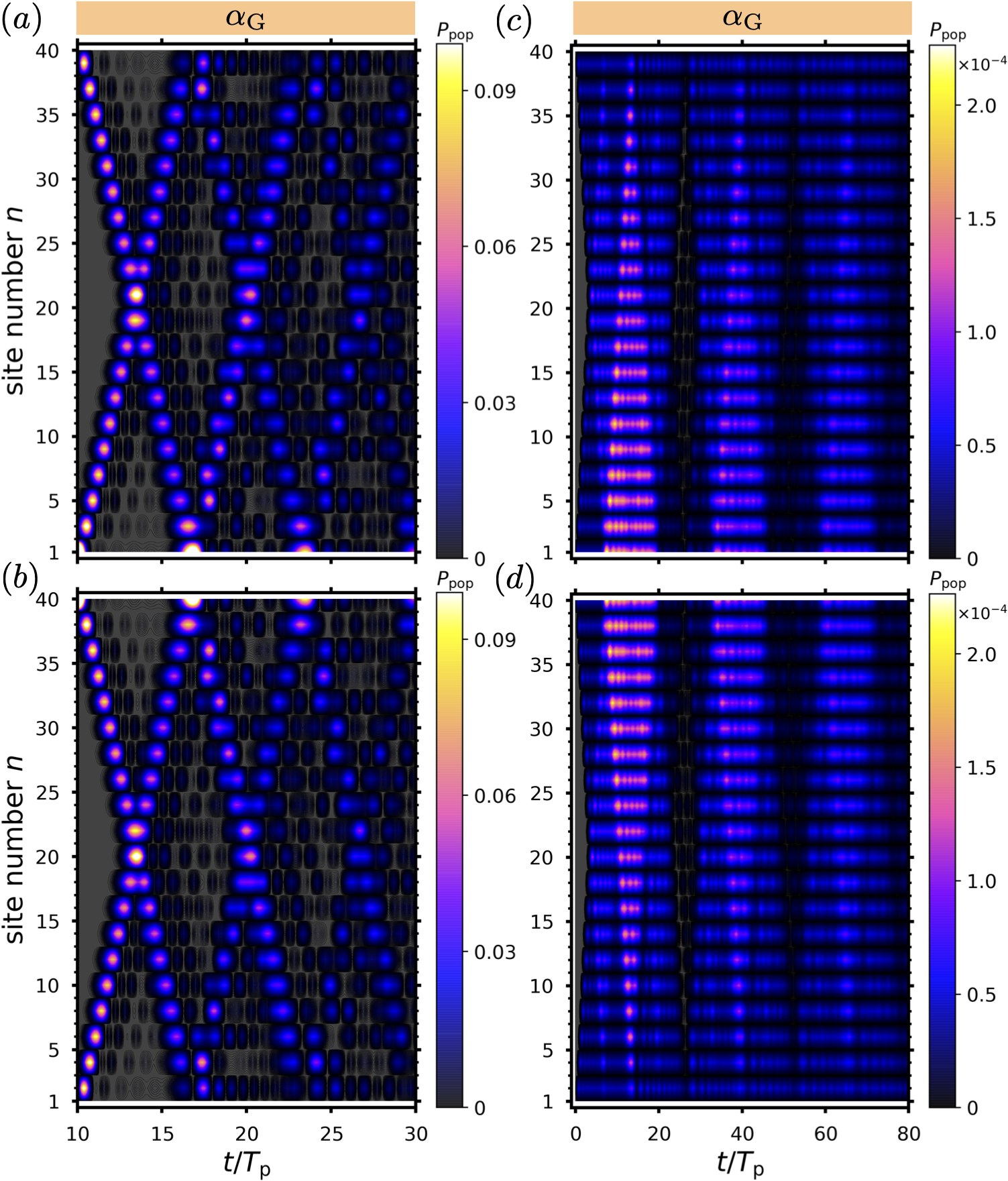}
\caption{
Occupation number $P_{\rm pop}$ at sublattice A with odd sites (a, c) and sublattice B with even sites (b, d) during the time period $10T_p-30T_p$. Panels (a, b) correspond to the Fig.~\ref{fig:multiparticle}(b) with the quantum quench scheme starting from the topological phase, while (c, d) correspond to Fig.~\ref{fig:multiparticle_B}(b) in the gapless phase without quench.}
\label{fig:multiparticle_sublattice}
\end{figure}

To better understand the 
difference between the topologically induced and conventional interferences of the gapless phase [Fig.~\ref{fig:multiparticle}(b) and Fig.~\ref{fig:multiparticle_B}(b), respectively], we show in Fig.~\ref{fig:multiparticle_sublattice} the site occupation probabilities on the separate sublattices A (odd sites) and B (even sites) [Eq.~(\ref{eq:hamiltonian})]. Here the left panels (a) and (b) show the sublattice site populations for the case of the quantum quench, and the right panels (c) and (d) show the corresponding populations in the case of no quench. 

\begin{figure}
\centering
  \includegraphics[width=.98\columnwidth]{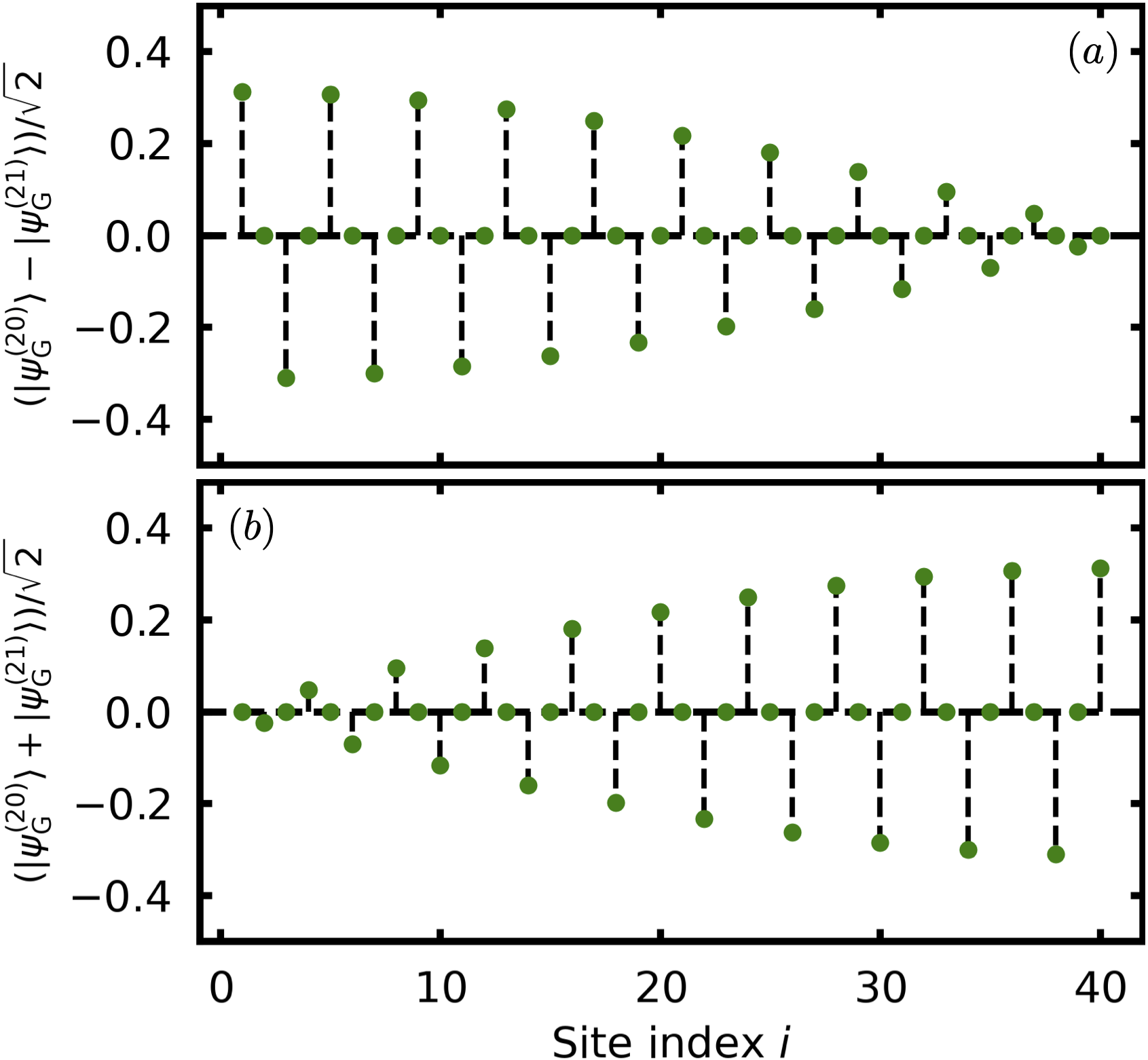}
\caption{
The superposition of highest filled and lowest unfilled eigenstates of the SSH model in the gaplesss phase by setting $\alpha$ to $\alpha_{\rm G}=0.5\pi$: (a) $(|\psi_{\rm G}^{(20)}\rangle-|\psi_{\rm G}^{(21)}\rangle)/\sqrt{2}$ and (b) $(|\psi_{\rm G}^{(20)}\rangle+|\psi_{\rm G}^{(21)}\rangle)/\sqrt{2}$, with corresponding eigenenergy in Fig.~\ref{fig:schematicSpectrAmplit}(b). Parameters are same as in Fig.~\ref{fig:schematicSpectrAmplit}(b). }
\label{fig:gapless_plus_minus_states}
\end{figure}

The occupation probabilities at odd and even sites of Fig.~\ref{fig:multiparticle}(b) under the quantum quench are represented in Fig.~\ref{fig:multiparticle_sublattice}(a) and Fig.~\ref{fig:multiparticle_sublattice} (b), respectively.
The odd- and even-site occupation probabilities of Fig.~\ref{fig:multiparticle_B}(b) for the gapless phase in the absence of any quench are shown in  Fig.~\ref{fig:multiparticle_sublattice}(c) and Fig.~\ref{fig:multiparticle_sublattice}(d), respectively.
It is evident that Fig.~\ref{fig:multiparticle_sublattice}(d) which involves only even sites is quite similar to  Fig.~\ref{fig:multiparticle_B}(d) where the whole lattice is considered and only the first site is subject to pumping. In Ref.~\onlinecite{Suppl} we show that Fig.~\ref{fig:multiparticle_sublattice}(c) is also similar to the corresponding plot with the last site pumped [see Fig.~S3(d)]. These similarities do not hold for the occupation probabilities under the quantum quench scheme, i.e., for comparison of Fig.~\ref{fig:multiparticle}(b) with Fig.~\ref{fig:multiparticle_sublattice}(a) and Fig.~\ref{fig:multiparticle_sublattice} (b).
In Fig.~\ref{fig:gapless_plus_minus_states} we plot the two real valued equal superpositions of the highest filled and lowest unfilled eigenstates in the gapless phase $\alpha_{\rm G}=0.5\pi$, i.e., $|\psi_{\rm G}^{(20)}\rangle$ and $|\psi_{\rm G}^{(21)}\rangle$, respectively. It is clear from this that the observed odd- or even-site occupations are accounted for by the odd superposition $(|\psi_{\rm G}^{(20)}\rangle-|\psi_{\rm G}^{(21)}\rangle)/\sqrt{2}$ [Fig.~\ref{fig:gapless_plus_minus_states}(a)] and by the even superposition $(|\psi_{\rm G}^{(20)}\rangle+|\psi_{\rm G}^{(21)}\rangle)/\sqrt{2}$ [Fig.~\ref{fig:gapless_plus_minus_states}(b)], respectively.

Calculations with the lattice maintained at the topologically trivial phase give rise to quite different interference patterns, shown in the Supplementary Material~\cite{Suppl}. In this situation weaker oscillating occupations are seen at bulk sites than at edge sites, i.e., quite different behaviour from not only the topological interference patterns but also from the non-quenched patterns seen for the nontrivial topological and gapless phases shown in Fig.~\ref{fig:multiparticle_B}. %
We have also analyzed the alternative dynamics of quenching from the topological to the trivial non-topological phases by setting $\alpha$ to $\alpha_{\rm tr}=0.25\pi$ and the resulting interference patterns (see Fig.~S11~\cite{Suppl}) look different from patterns from the above main quench scheme considered in this work.

In summary, we observe two types of topological quantum interference under pumping of the end sites of the finite SSH lattice. The first is seen in the gapless phase after quenching from the topological phase. Here the interference starts to develop from the large component of pre-quench topological edge excitations present in time evolved state at the quench time. 
The second interference pattern is seen after switching back to the topological phase and is quite distinct in form, as well as being accompanied by the expected high-density of edge excitations due to the pumping at edge sites.
The detailed analysis of this section has shown that these interference phenomena 
are quite distinct from conventional quantum interference behaviors seen in either the topologically nontrivial or gapless phases, or from quenching from the topological into the trivial phase.

\begin{figure}
\centering
  \includegraphics[width=.96\columnwidth]{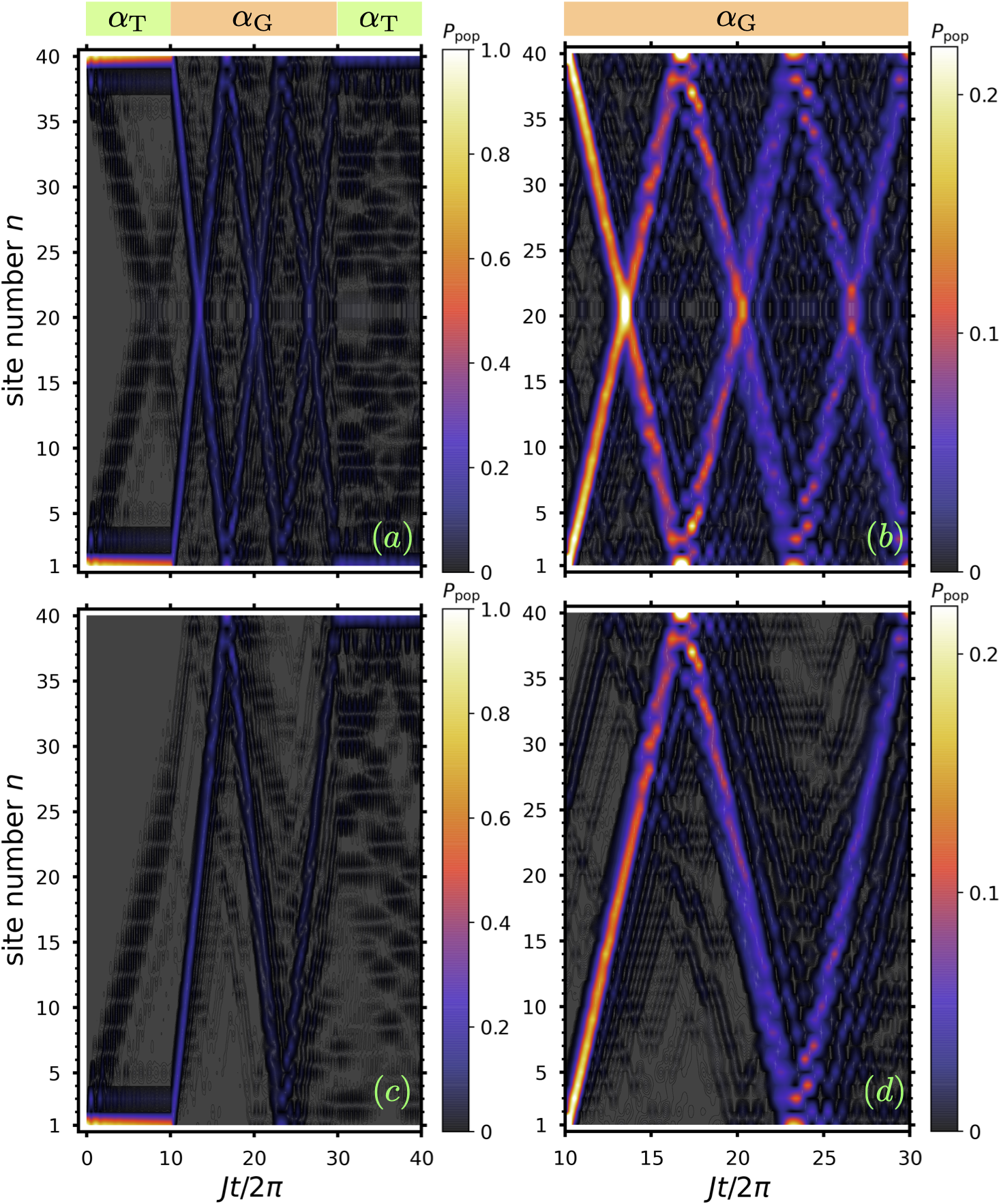}
\caption{
Time-dependent occupation number $P_{\rm pop}$ in the case of two excitations (top panels) and of one excitation (bottom panels) inserted in the non-pumped SSH lattice at $t=0$.
Panels (a) - (b): the two excitations initially occupy both ends ($A_{1}(0)=B_{20}(0)=1$). Panels (c) - (d): one excitation initially occupies the first site ($A_{1}(0)=1$, $B_{20}(0)=0$).
Topological interference is observed in (a) with $\alpha_{\rm T}=0.75\pi$, $\alpha_{\rm G}=0.5\pi$, and the switching time $Jt_a/2\pi=10$, $Jt_b/2\pi=30$. 
Panels (b) and (d) are zoom-in views of the patterns during $10<Jt/2\pi<30$ in (a) and (c), respectively.
Here the pumping fields are switched off ($F_A=F_B=0$), $\gamma=0.0025$, and all other parameters are the same as those in Fig.~\ref{fig:multiparticle}. 
All energy quantities are defined in units of $J(=\omega_p)$ (e.g., $J_1=J_2=0.5J$ at $\alpha_{\rm G}=0.5\pi$) if not specified. 
}
\label{fig:twoparticle}
\end{figure}

\section{Two-excitation quantum interference in the absence of pumping}
\label{sec:double_nopump}

To further elucidate the basic features of quantum interference between topological states, we now consider the case of two excitations initially occupying the edge sites, with both pumping fields turned off (i.e., $F_A=F_B=0$). We run the calculations with excitations present at $t=0$, using the same switching protocol as above. 
Explicitly, we solve Eq.~(\ref{eq:equation}) with Eq.~(\ref{eq:scheme}) and initial condition $A_{1}(0)=B_{N}(0)=1$ or $A_{1}(0)=1$, $B_{N}(0)=0$ for the case of two excitations or one excitation, respectively.
Fig.~\ref{fig:twoparticle}(a) shows how the two excitations evolve on the lattice, with a zoom-in view of the interference pattern shown in Fig.~\ref{fig:twoparticle}(b).
It is evident that the interference pattern in Fig.~\ref{fig:twoparticle}(b) is similar to that in Fig.~\ref{fig:multiparticle}(b), implying  
robustness of the topological interference pattern against the number of excitations.
However due to the effect of dissipation, this two-excitation interference pattern becomes less visible with the increase of time. 
Compared with the pumped case in Fig.~\ref{fig:multiparticle}(a), there is a noticeable feature of a small wave-package moving from the edge to the bulk during the initial times (i.e., before the first quench) in Fig.~\ref{fig:twoparticle}(a). This is because the initial occupation at edge sites in the absence of pumping fields is ready to start to decrease a little bit into bulk, while in the pumped case it needs time for the pumping fields to build up the occupation at the edge sites, which is evidenced by a close comparison between Figs.~\ref{fig:twoparticle}(a) and ~\ref{fig:multiparticle}(a).

The behavior of the two edge states that participate in the interference is further revealed by considering a single excitation initially at the first or last site.  Fig.~\ref{fig:twoparticle}(c) shows the results for a single excitation initially at the first site, with a zoom-in view shown in Fig.~\ref{fig:twoparticle}(d). 
In these cases we also see that it is the delocalization of excitations from the edge to bulk, rather than smaller amplitude of pre-quench bulk excitations in the topological phase, that nucleate the interference in the gapless phase after the quench.
Similar behavior is found when the single excitation is initially localized at the last site (see~\cite{Suppl}), consistent with the symmetry of the $2N =40$ chain.

\begin{figure}
\centering
  \includegraphics[width=1.0\columnwidth]{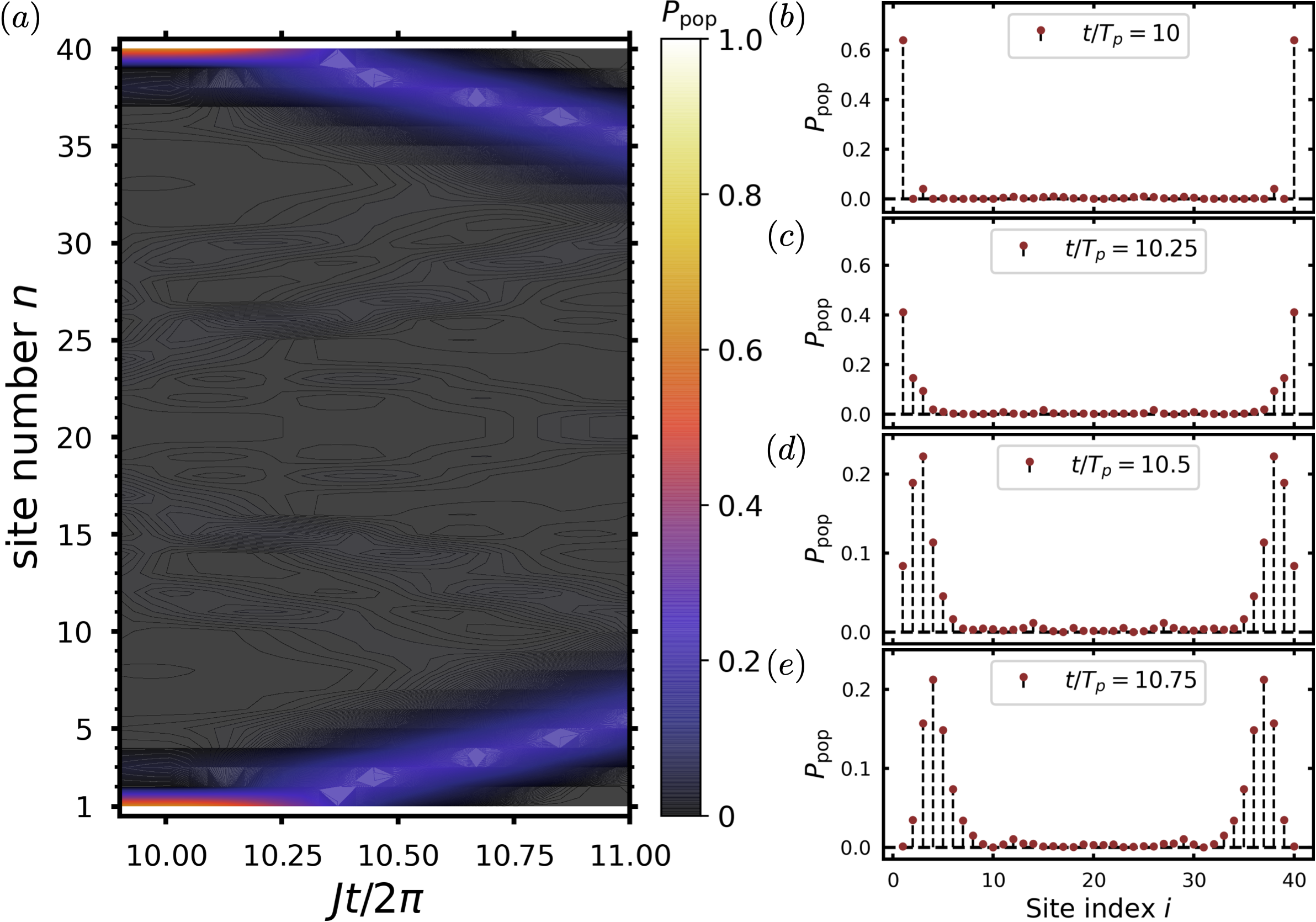}
\caption{
(a) A zoom-in view of Fig.~\ref{fig:twoparticle}(a) around the first switch at $t/T_p=10$. The occupation number $P_{\rm pop}$ across the SSH lattice when (b) $Jt/2\pi=10$, (c) $Jt/2\pi=10.25$, (d) $Jt/2\pi=10.5$, and (e) $Jt/2\pi=10.75$.
}
\label{fig:twoparticle_quenchHysteresis}
\end{figure}

\begin{figure}
\centering
  \includegraphics[width=.98\columnwidth]{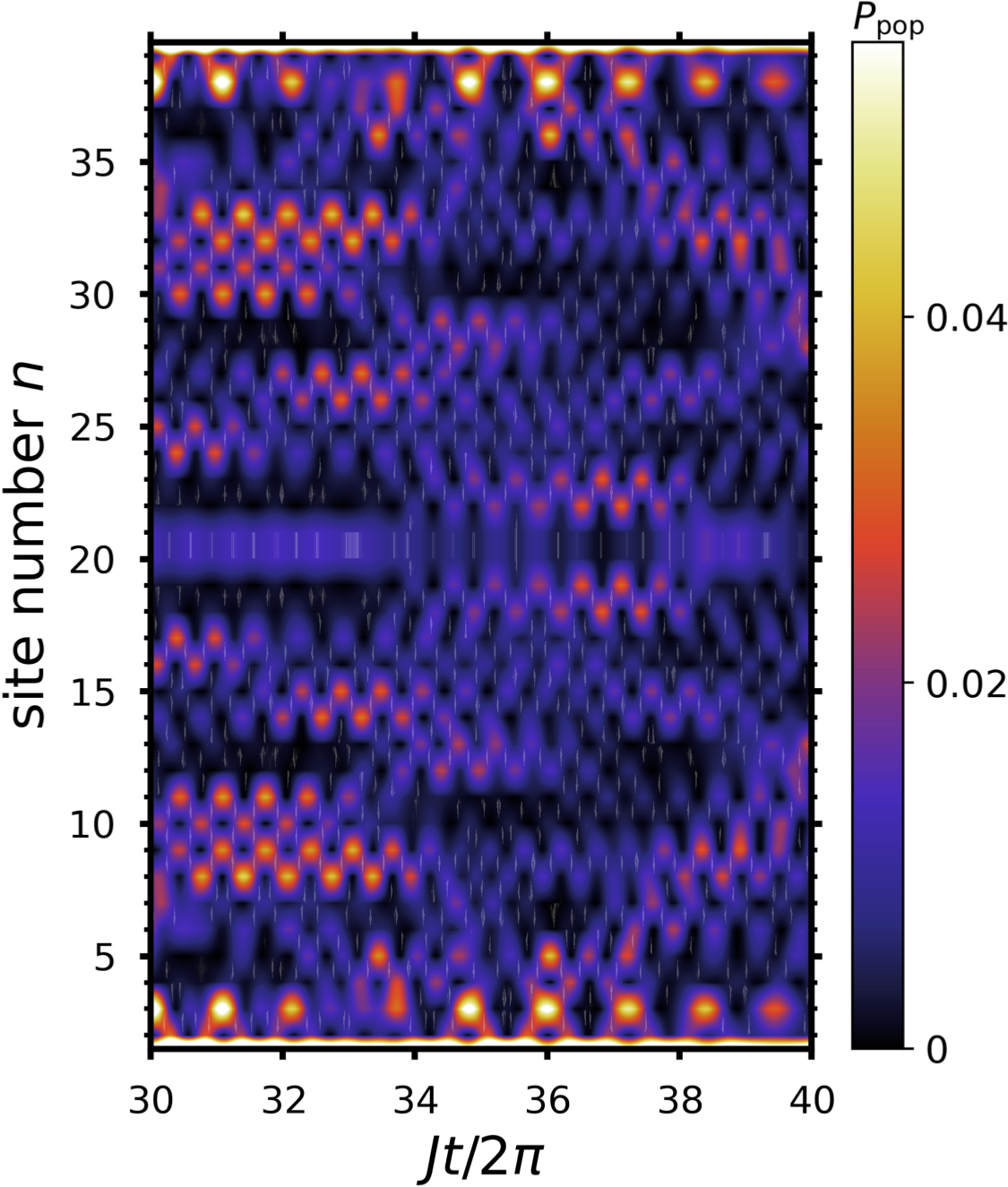}
\caption{A zoom-in view of the interference pattern in the topological phase quench period $30<Jt/2\pi\le40$ of Fig.~\ref{fig:twoparticle}(a). 
The plot shows the occupation number $P_{\rm pop}$ in the case of two excitations initially occupying both ends, i.e., $A_{1}(0)=B_{20}(0)=1$ at $t=0$, with no pumping fields ($F_A=F_B=0$).
Other parameters are same as in Fig.~\ref{fig:twoparticle}(a), i.e., 
$\alpha_{\rm T}=0.75\pi$, $\alpha_{\rm G}=0.5\pi$, $Jt_a/2\pi=10$, $Jt_b/2\pi=30$ 
$\gamma=0.0025$, $N=20$, $\varepsilon=1$, $V_0=0.125$, and $\mu/k^2=0.25$.}
\label{fig:twoparticle_A_plus}
\end{figure}

Around the first switch at $t=10T_P$, we also observe the phenomenon of hysteresis where the delocalization of the edge-state population into the bulk lags behind the switch, as demonstrated in both Fig.~\ref{fig:twoparticle_quenchHysteresis} (a zoom-in view of Fig.~\ref{fig:twoparticle}(a)) and Fig.~\ref{fig:twoparticle_quenchHysteresis}(b-e) for the lattice occupation number at e.g., $t=10T_p$ (immediately before the quench), and $t=10.25T_p, 10.5T_p, 10.75 T_p$ (after the quench), respectively.
After switching back to the topological phase in Fig.~\ref{fig:twoparticle}(a), we again observe an interference in the bulk in addition to regenerated excitations at the edges. This is clearly revealed by the zoom-in view presented in Fig.~\ref{fig:twoparticle_A_plus} which shows similar features to that of Fig.~\ref{fig:multiparticle_A_plus}. 
To improve the visibility of the pattern in the bulk sites in Fig.~\ref{fig:twoparticle_A_plus} we include here all sites from $n=2$ to $39$ and change the scaling of the color bar from that of Fig.~\ref{fig:twoparticle}(a).

\begin{figure}
\centering
  \includegraphics[width=.96\columnwidth]{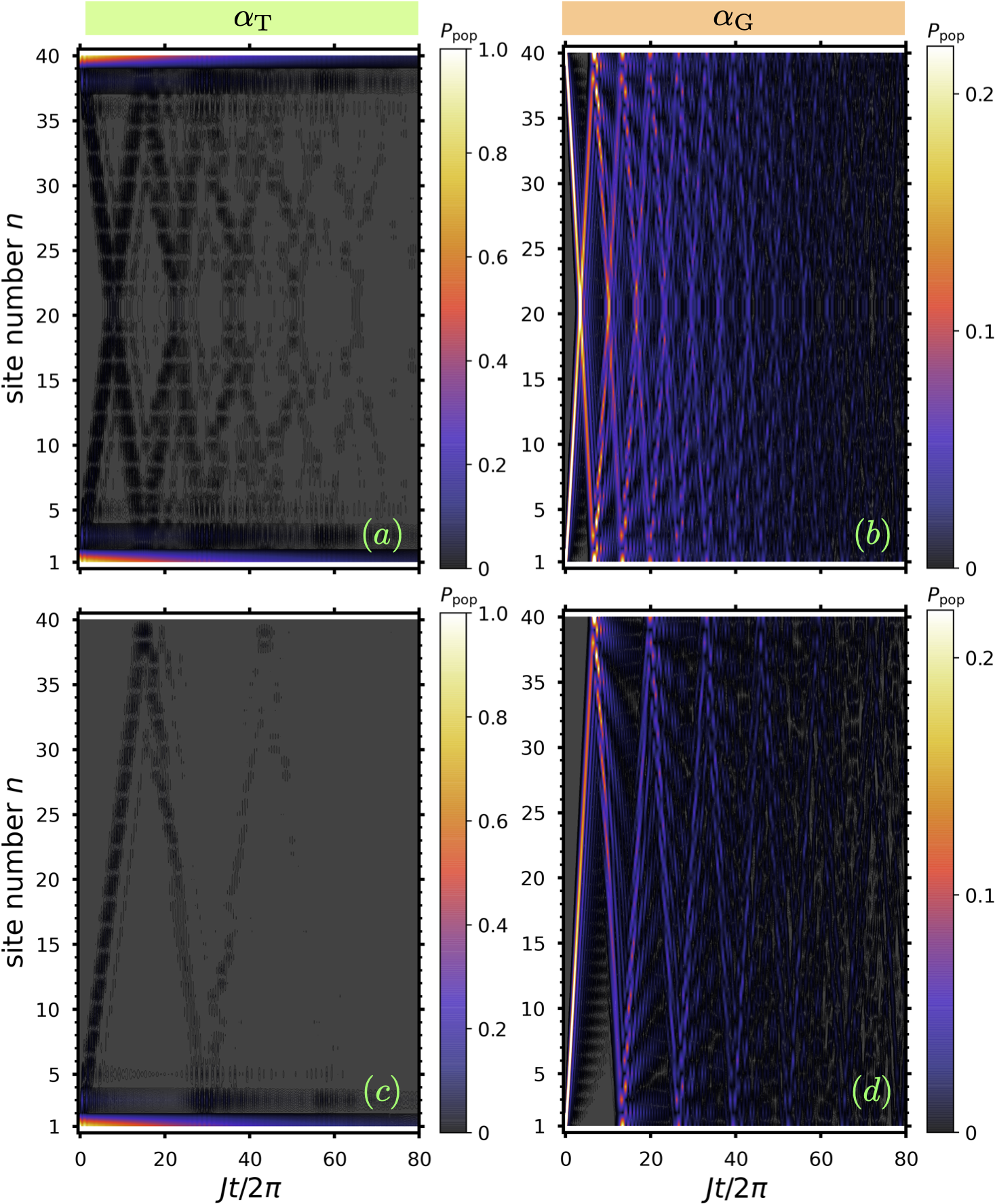}
\caption{
Time-dependent occupation number $P_{\rm pop}$ in the case of two excitations (top panel) and one excitation (bottom panels) inserted in the non-pumped SSH lattice at $t=0$, for the topologically nontrivial phase (left panels, $\alpha_{\rm T}=0.75\pi$) and for the gapless phase (right panels, $\alpha_{\rm G}=0.5\pi$).
Panels (a)-(b): the two excitations initially occupy both ends ($A_{1}(0)=B_{20}(0)=1$). Panels (c)-(d): one excitation occupies the first site ($A_{1}(0)=1$, $B_{20}(0)=0$).
Here the pumping fields are switched off ($F_A=F_B=0$), $\gamma=0.0025$, and all other parameters are same as those in Fig.~\ref{fig:twoparticle}.}
\label{fig:twoparticle_B}
\end{figure}

In this case we also consider the dynamics in the topologically nontrivial ($\alpha_{\rm T}=0.75\pi$) and gapless ($\alpha_{\rm G}=0.5\pi$) phases separately, i.e., without switching, as was done above in the case of the pumped many excitation dynamics.
Figure~\ref{fig:twoparticle_B}(a) shows that when the dynamics are confined to  the topological phase (here $\alpha_{\rm T}=0.75\pi$), the topological states stay localized at both ends and no interference is evident.  Figure~\ref{fig:twoparticle_B}(c) shows the corresponding dynamics when a single excitation is initially localized at the first site. Here while the excitation is mainly localized at the edges, an additional weak oscillation is observed, which is not noticeable in the pumped many excitation case in Fig.~\ref{fig:multiparticle_B}(c).
In the gapless phase,  Fig.~\ref{fig:twoparticle_B}(b), in the absence of switching we nevertheless find an interference between non-topological states. This pattern is quite different from the interference patterns seen in either the pumped single-excitation case [Fig.~\ref{fig:twoparticle_B}(d)]  or the pumped topological interference case [Fig.~\ref{fig:twoparticle}(b)].
It is also quite unlike our findings for the corresponding case of pumped many excitations in the gapless phase in Figs.~\ref{fig:multiparticle_B}(b)-(d) and Fig.~\ref{fig:multiparticle}(b), indicating a clear dependence of the interference pattern of the non-quenched gapless phase on the number of the excitations in the lattice.

\begin{figure*}
\centering
  \includegraphics[width=1.4\columnwidth]{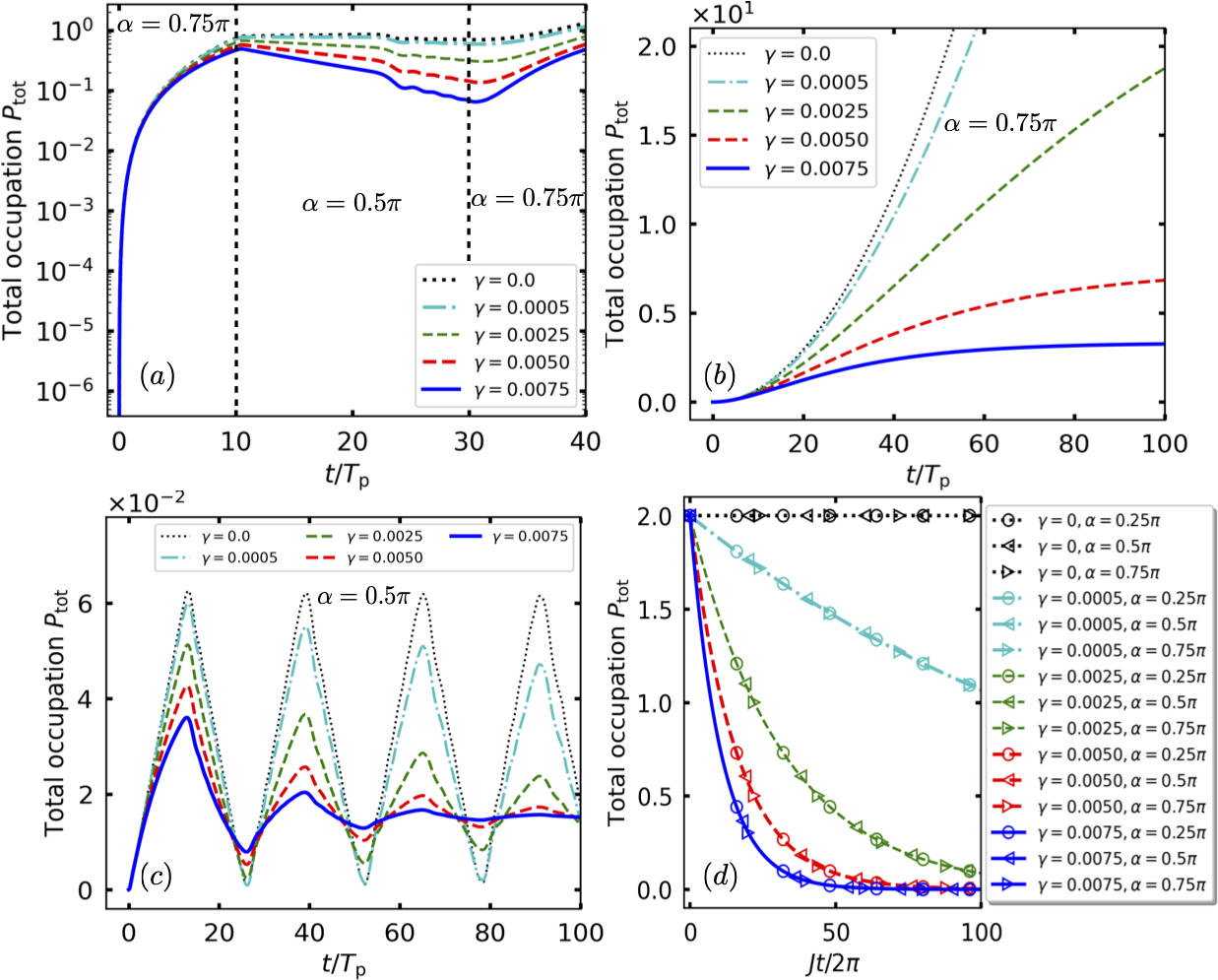} 
\caption{
Evolution of the total number $P_{\rm tot}$ of excitations  for the pumped many-excitation case [panels (a), (b), (c)] and non-pumped two-excitation case [panel (d)] for various values of damping rate $\gamma$. Panels (a) interference scheme, (b) topological nontrivial phase ($\alpha=0.75\pi$), and (c) gapless phase ($\alpha=0.5\pi$) corresponding to Fig.~\ref{fig:multiparticle}(a), Fig.~\ref{fig:multiparticle_B}(a), and Fig.~\ref{fig:multiparticle_B}(b), respectively. In panel (d), the dotted, dash-dotted, thin-dashed, thick-dashed, and solid curves correspond to $\gamma=0, 0.0005, 0.0025, 0.005, 0.0075$ respectively, while the circle, left and right triangles are for $\alpha=0.25\pi, 0.5\pi, 0.75\pi$, respectively.
Other parameters for (a), (b, c) and (d) are the same as those in Figs.~\ref{fig:multiparticle}, ~\ref{fig:multiparticle_B}, and~\ref{fig:twoparticle}, respectively.}
\label{fig:totalpopu}
\end{figure*}

\section{Total excitation population}
\label{sec:total+dissipation}

We now consider the total excitation population  of the entire lattice as a function of the damping rate $\gamma$. 
We employ here the non-Hermitian Hamiltonian formulation employed in Monte Carlo wave function dynamics~\cite{PlenioKnight98rmp} that describes the conditional dynamics of the density operator within a Lindblad equation of motion, under the assumption that no photons from spontaneous emission have been detected~\cite{LiSarovarWhaley21njp}.

Fig.~\ref{fig:totalpopu}(a) shows the total population that corresponds to the many-excitation interference in Fig.~\ref{fig:multiparticle}(a).
For a given value of $\gamma$, it is evident that the total population first increases before the first switch, $t\le t_a=10T_p$, after which it then exhibits small fluctuations during the interference period between the second switch, $t_a<t\le t_b=30T_p$. After the second switch, i.e., $t_b<t$, the excitation population further increases.
These behaviors can be understood as follows. When the lattice is in the topological phase with $\alpha_{\rm T}=0.75\pi$, the pumping-induced excitations are mainly localized at the two edges of the lattice and the total population can increase rapidly, even in the presence of damping at the edge sites.
However on switching to the gapless phase, the excitations spread out to the entire lattice and damping now occurs at all sites, rather than just at the two end sites. This can slow down or even oppose the increase [see, e.g., the cyan ($\gamma=0.0005$) or red ($\gamma=0.005$) curves respectively in Fig.~\ref{fig:totalpopu}(a)], depending on how large the damping effect is. 

The effect of varying $\gamma$ becomes more obvious when considering each of the three quantum phases individually, i.e., the topologically nontrivial phase, the gapless phase, and the trivial phase.
In the topological nontrivial phase, Fig.~\ref{fig:totalpopu}(b) shows that the total population increases monotonically for any value of $\gamma$. In the gapless phase, Fig.~\ref{fig:totalpopu}(c) shows that the total population first oscillates and then approaches a steady value, while 
in the trivial phase it oscillates even more rapidly (see the supplementary information~\cite{Suppl}).
Furthermore, in the gapless phase, Fig.~\ref{fig:totalpopu}(c) shows not only the oscillation period but also the steady state occupation are independent of the value of $\gamma$ when it is nonzero. This reflects the presence of a non-equilibrium steady state due to a balance between pumping and dissipation.
We also see that decreasing $\gamma$ enhances oscillations in the gapless phase [Fig.~\ref{fig:totalpopu}(c)] (as well as in trivial phase [Fig.~S5(d)]~\cite{Suppl}) and accelerates the increase of the total population in the topological nontrivial phase [Fig.~\ref{fig:totalpopu}(b)].
In other words, the total population in both the gapless and topological nontrivial phases is suppressed by increasing the damping rate, and the interference of topological states, represented by the occupation states of excitations on lattice sites, is correspondingly reduced. 
This is confirmed by the calculations in Fig.~\ref{fig:multiparticle} which are carried out with different values of the damping rate.
Finally, on comparing the behavior for two pumping fields [see Fig.~\ref{fig:totalpopu}(b), Fig.~\ref{fig:totalpopu}(c), or Fig.~S5(d)~\cite{Suppl} for the topological nontrivial, gapless, or trivial phases, respectively] with that for a single pumping field applied at the first or last site (additional calculations not shown here), we see that two pumping fields double the total population in each phase.
Importantly, it is clear that this doubling behavior does not apply for the occupation at individual sites due to quantum interference. This is evident from comparison of Fig.~\ref{fig:multiparticle}(b) with Fig.~\ref{fig:multiparticle}(d) and Fig.~S3(b)~\cite{Suppl} or of Fig.~\ref{fig:twoparticle}(b) with Fig.~\ref{fig:twoparticle}(d) and Fig.~S4(b)~\cite{Suppl}.

For the case of two initial excitations without pumping, the total population in the lattice does not change when $\gamma=0$. This is shown in Fig.~\ref{fig:totalpopu}(d) (compare open 
symbols).
For a given value of the optical lattice phase tuning $\alpha$, increasing $\gamma$ from $0.0005$ to $0.0075$ as indicated by the cyan and blue curves respectively, does accelerate the decay of the total population, as expected.
However, for a given value of $\gamma$, for certain values of $\alpha$ we can see the same decay behavior of the total population in all three  phases. This is illustrated in Fig.~\ref{fig:totalpopu}(d) for the phase tuning parameter values $\alpha=0.25\pi,0.5\pi,0.75\pi$ representing topological trivial, gapless, and nontrivial phases, respectively.
 
This {\it $\alpha$-independence} 
is distinct from the decay causing suppressed interference of topological states (verified by an extra calculation of Fig.~\ref{fig:twoparticle} but with  a different value of damping rate~\cite{Suppl}). 
Here the $\alpha$-independence
indicates instead that the topology does not play a role, which is in sharp contrast to what was seen for pumped many-excitation case shown in Fig.~\ref{fig:totalpopu} (b) and (c).
To understand this difference between the pumped many-excitation and unpumped two-excitation situations, we note that, while the pumping fields in Fig.~\ref{fig:totalpopu} (b) and (c) produced steadily increasing excitations at the lattice edges, the dynamics for two initial excitations presented in Fig.~\ref{fig:totalpopu}(d) always have a fixed number of excitations which then decrease universally at each occupying sites with identical damping rate. This universal decrease is independent of the quantum phase of the lattice.
In Ref.~\onlinecite{Suppl} we show that for the case of two excitations, such a topology-independence damping is also found in the presence of phase switching (Fig.~S8).  This results in  
quite different time dependence of total excitation population $P_{\rm tot}$ from that seen in Fig.~\ref{fig:totalpopu}(a) for the case of pumped many-particle excitations in the presence of phase switching. %

In summary, when the possibility of spontaneous emission from the excitonic states is included, analyzing the conditional dynamics within a Lindblad description of the interaction of the excitonic states with a radiative bath yields a  non-Hermitian SSH system Hamiltonian with a local dissipative term at each lattice site.  The results above show that this can suppress the total excited state population of the lattice, unlike well-known situations where topological states are protected against local perturbations. In particular, for the case of  two initial excitations without pumping, we  find that the decay of the total population does not depend on whether the system is in the topological phase or not (i.e., no dependence on the value of $\alpha$). Thus in this case the overall excitation decay rate is independent of the quench protocol.  This is  in sharp contrast to the case of the pumped non-Hermitian SSH system where we find the total population shows sharp transitions on switching, although there is again no protection against decay of excitations.

\section{Discussion and conclusions}

In this work we have presented a realization of topological quantum interference in a pumped SSH lattice represented by the excitation occupation number on each lattice site. The lattice is assumed to be tunable, as in recent experiments with trapped atoms. 

We have investigated both pumped many-excitation and unpumped two-excitation quantum interference induced by either topological or non-topological states. 
Our results show that similar interference patterns  exist for pumped many- and non-pumped two-excitation cases and this similarity could be fundamentally due to the fact that both of them belong to a single-particle interference. 
Topological interference patterns are distinguishable from those of non-topological states. The effects of dissipation on both the interference patterns and the total excitation population of the whole lattice have been studied. For a system initialized with  two excitations at each end of the lattice, we additionally find that the decay of the total population does not depend on whether the system is in topological phase or not, which is in sharp contrast to the many excitation case.

It is interesting to compare the current results with the interferences observed in the recent  experiment based on a photonics realization of the off-diagonal Harper model~\cite{Tambasco18SciAdv}. 
Our work shows topological interference patterns that are quite distinct from the interference patterns of non-topological states.
We note that the basic idea of Ref.~\onlinecite{Tambasco18SciAdv}, to bring two initial edge states into the bulk and let the resulting states interfere, has some similarity with the evolution of pre-quench topological edge states into the bulk seen in the current work. The bulk gap in both cases is either closed or reduced in order to let the two topological states meet. 
However the quench protocol employed here is very different from the adiabatic delocalization from  edges to the bulk that was employed in Ref.~\onlinecite{Tambasco18SciAdv}.  
Furthermore, before interference the evolution from initial edge states to the pre-quench bulk states (more noticeable in the non-pumped case) in the current study indicates the pre-interference states are not exactly from the delocalization of the pre-quench topological edge states. 
We expect that there exist other possible schemes to generate topological entanglement, that might be realizable in other quantum simulation platforms.

Interestingly, our work also reveals that the total population of the SSH latice in the topological phase ($\alpha=0.75\pi$) could be suppressed by an increase of the damping rate (non-Hermitian term in the Hamiltonian Eq.~(\ref{eq:hamiltonian}) for conditional evolution with no spontaneous emission detections) [see Fig.~\ref{fig:totalpopu}(a) or \ref{fig:totalpopu}(b) and Fig.~\ref{fig:totalpopu}(d) for many- and two-excitation cases, respectively]. In particular the interference of topological states presented in Fig.~\ref{fig:multiparticle} or \ref{fig:twoparticle} is correspondingly reduced, in contrast to the widely known robustness of topological states against local perturbations.

Given the recent experimental demonstrations of SSH lattices~\cite{Schomerus13ol,WeimannSzameit17nmat,AtalaDemlerBloch13nphys,Tambasco18SciAdv,de2019observation} our proposal of topological quantum interference based on a pumped SSH model with realistic damping appears timely and achievable with current technology.
In the current work, to clearly demonstrate interference features we have considered edge-site pumping. Other approaches that we expect would give similar interferences include  bulk excitation~\cite{KrausZilberberg12prl,Quandt12} and
high-frequency modulation of the SSH lattice~\cite{Krivosenko18pra,GomezPlatero13prl}.

It will be interesting in future work to consider what the effect of excitonic interactions might have on the topological quantum interferences observed in the current study. We expect that an effective mean-field description as implicitly assumed here for the pumped SSH lattice with multiple excitations 
would become increasingly accurate in the limit of large excitation numbers for finite range couplings. It would also be expected to be increasingly accurate as the excitonic interaction strength decreases. 
Going beyond the current effective description to consider the effects of excitonic interactions requires analysis within the larger many-body Hilbert space.  Clearly the strength and nature of the excitonic interactions will play a critical role in modifying the dynamics seen in this work. Thus one might expect that the density of excitations at each site will decrease with a repulsive excitonic interaction. 
Future work could usefully explore how the interplay between an excitonic interaction energy and the hopping amplitudes influences the topological quantum interference.

In this work we have studied the evolution of excitations both with and without continuous pumping actions. The resulting interference patterns are similar, indicating their existence is independent of the means of generation. However the interferences are less clearly visible in the absence of the pumping fields, due to the unmitigated decrease of the excitation populations due to damping in this situation.
Reservoir engineering techniques to decrease the decay rate $\gamma$ may provide routes to preservation of the interference of topological states~\cite{Verstraete09nphys,MullerZoller12,Vuglar18prl}. 
In the present calculations we 
have assumed identical dissipation due to e.g., spontaneous emission for each lattice site with the same onsite energy. This assumption is reasonable when the typical scale 
of the relevant coupled environmental degrees of freedom is larger than the lattice size. 
It will be of interest to further investigate the effects of decoherence, including both dephasing and dissipative terms from different environmental degrees of freedom, to explore the extent to which  the interference of topological states survives in a lattice that is subject to dynamical disorder, as well as in an inhomogeneous lattice subject to static disorder.

Another interesting direction for further work is to use techniques such as pump-probe spectroscopy to measure additional properties such as phase-modulated nonlinear spectra, which may provide useful information in the case of weak interactions~\cite{gessner2014nonlinear,Bruder15pra,LiEisfeld17pra}.
Going beyond the resonant pumping and semiclassical treatment in the present work, a useful additional direction for further work includes investigation of the effects of off-resonant rather than resonant pumping fields, as well as interaction of the SSH lattice with a quantum environment,

Finally we point out that the model in this work becomes the previously studied non-Hermitian ${\cal PT}$-symmetric SSH model~\cite{Schomerus13ol,Lieu18prb} when considering gain and loss at all sites (parameterized by $\gamma_{2j-1}=-\gamma_{2j}=-\gamma$) instead of the situation of coherent pumping fields at end sites and identical dissipation at all sites considered in this work. 
Single edge-mode lasing has been proposed and experimentally observed in an SSH array constructed on a hybrid silicon platform~\cite{ZhaoSchomerusFeng18ncomms} and with microring resonators~\cite{PartoKhajavikhan18prl}. 
While the physical realization of non-Hermitian ${\cal PT}$-symmetric model in quantum systems is generally limited by noise introduced via incoherent gain and loss processes~\cite{ScheelSzameit18epl}, it has been shown that a dissipation-free Hermitian quantum Hamiltonian can nevertheless produce an effective non-Hermitian SSH model~\cite{WangClerk19pra}.
The recent experimental demonstration of generation of a non-Hermitian superconducting qubit by postselection on a three-level transmon circuit also offers an alternative primitive for construction of extended non-Hermitian Hamiltonian systems \cite{NaghilooMurch19nphys}.

\section{Methods \label{sec:method}}

The semiclassical dynamics of the SSH lattice with dissipative decay and in the absence of pumping fields is governed by the time-dependent Schr\"{o}dinger equation,
\begin{eqnarray}
i\partial_t |\psi(t)\rangle &=& H_{\rm SSH} |\psi(t)\rangle , \label{eq:schrodinger}
\end{eqnarray}
where $H_{\rm SSH}$ is given in Eq.~(\ref{eq:hamiltonian}). The wavefunction can be written as
\begin{eqnarray}
|\psi(t)\rangle &=& \sum_l \otimes \binom{A_l(t)}{B_l(t)} = \sum_l A_l|l,A\rangle +B_l|l,B\rangle ,
\end{eqnarray}
with time-dependent amplitudes $A_l$ and $B_l$. By inserting this into Eq.~(\ref{eq:schrodinger}) and including the pumping terms $F_A e^{i(\phi_{0A}-\omega_{pA} t)}$ and $F_B e^{i(\phi_{0B}-\omega_{pB} t)}$ as well, we obtain the evolution equations
\begin{eqnarray}
i\partial_t A_l &=& J_1 B_l + J_2 B_{l-1} + (\varepsilon -i\gamma) A_l +\delta_{l,1} F_A e^{i(\phi_{0A}-\omega_{pA} t)}, \notag\\
i\partial_t B_l &=& J_1 A_l + J_2 A_{l+1} + (\varepsilon -i\gamma) B_l +\delta_{l,N} F_B e^{i(\phi_{0B}-\omega_{pB} t)}. \notag
\end{eqnarray}
These equations are solved numerically, with boundary conditions $A_0=B_0=A_{N+1}=B_{N+1}=0$. For the case of many excitations, the initial conditions are $A_l(0)=B_l(0)=0$ with $F_A=F_B=0.01$ for two pumping fields and $F_A=1, F_B=0$ for a single pumping field.
In the case of dynamics with no pumping fields and initial double or single excitations, these initial conditions become $A_1(0)=B_{N}(0)=1$ for double initial excitation and $A_1(0)=1, B_{N}(0)=0$ (others are always zeros) with $F_A=F_B=0$ for single initial excitation.
All calculations in this work are made for resonant pumping, with $\omega_{pA}=\omega_{pB}=\omega_{p}=\varepsilon$,  $F_A=F_B=F$, and $\phi_{0A}=\phi_{0B}=\phi_{0}$.

{\bf Acknowledgements.}
We thank Robert Cook and Torin Stetina for valuable discussions and the anonymous referees for constructive criticisms that helped further improve this paper. 
Z.Z.L thanks C. H. Lam and J. Q. You for helpful comments and suggestions on the earlier version of this work. 
We acknowledge support from the U.S. Department of Energy, Office of Science under Contract No. DE-AC02-05CH11231.

\end{document}


\title{Supplemental material for ``Topological quantum interference in a pumped Su-Schrieffer-Heeger lattice"}

\author{Zeng-Zhao Li,$^{\dagger,*}$ Juan Atalaya,$^{\dagger}$ and K. Birgitta Whaley$^{\dagger,*}$}
\affiliation{$^{\dagger}$Department of Chemistry, University of California, Berkeley, California 94720, USA}
\affiliation{$^{*}$Berkeley Center for Quantum Information and Computation, Berkeley, California 94720, USA}

\begin{abstract}
We provide here a derivation of the hopping amplitudes and present additional analysis of eigenstate distributions, entanglement entropy dynamics, the behavior under pumping at the last site, damping effects, as well as an alternative quantum quench protocol. 
\end{abstract}

\maketitle

\section{Deriving the hopping amplitudes of Eq.~(2) in the main text}

%
The optical potential 
\begin{eqnarray}
V_{OL}(x,\tau) &=& V_0 |e^{i(kx+\alpha)} + e^{-ikx} + e^{i3kx}|^2
= V_0 [3+4\cos(2kx)\cos(\alpha(\tau))+2\cos(4kx)], \label{eq:V_OL}
\end{eqnarray}
can be generated by three laser fields with amplitudes proportional to $e^{i(kx+\alpha)}$, $e^{-ikx}$, and $e^{i3kx}$.  
We calculate the hopping amplitudes $\Delta_1 (\alpha), \Delta_2 (\alpha)$ by first solving wave functions of an approximate potential around the local minima of $V_{OL}(x,\tau)$ and using these to evaluate the relevant matrix elements.

\subsubsection{Local minima} 
The condition for local minima of the optical potential is
\begin{eqnarray}
\frac{\partial V_{OL}(x,\tau)}{\partial t} &=& 0 .
\end{eqnarray}
For Eq.~(\ref{eq:V_OL}), this gives
\begin{eqnarray}
\frac{\partial V_{OL}(x,\tau)}{\partial t}
&=& -8kV_0[\sin(2kx)\cos(\alpha(\tau)) +\sin(4kx)] ,
\end{eqnarray}
and the condition for local minima reduces to  
\begin{eqnarray}
 \cos(2kx) = -\frac{\cos(\alpha(\tau))}{2}. \label{eq:minima_relation}
\end{eqnarray}
This has two possible sets of specific solutions:
\begin{eqnarray}
\clubsuit:
\cos(\pi+2kx) &=& \frac{\cos(\alpha(\tau))}{2} \rightarrow 2kx = -\pi + \arccos\big[\frac{\cos(\alpha(\tau))}{2}\big], \notag\\
\spadesuit: 
\cos(-\pi-2kx) &=& \frac{\cos(\alpha(\tau))}{2} \rightarrow 2kx = -\pi - \arccos\big[\frac{\cos(\alpha(\tau))}{2}\big], \notag
\end{eqnarray}
or
\begin{eqnarray}
\heartsuit:
\cos(\pi-2kx) &=& \frac{\cos(\alpha(\tau))}{2} \rightarrow 2kx = \pi - \arccos\big[\frac{\cos(\alpha(\tau))}{2}\big], \notag\\
\diamondsuit:  
\cos(-\pi+2kx) &=& \frac{\cos(\alpha(\tau))}{2} \rightarrow 2kx = \pi + \arccos\big[\frac{\cos(\alpha(\tau))}{2}\big]. \notag
\end{eqnarray}
One can choose either the set $\clubsuit, \spadesuit$ or the set $\heartsuit, \diamondsuit$, with the two sets related to each other via a period of $2\pi$.  
By adding the period $2n\pi$ into the specific solutions, i.e., $ \pi \pm \arccos\big[\frac{\cos(\alpha(\tau))}{2}\big]$, we then obtain the general solution for local minima of Eq.~(\ref{eq:V_OL}), namely,
\begin{eqnarray}
2kx_n^+ &=& \pi + \arccos\big[\frac{\cos(\alpha(\tau))}{2}\big] + 2n\pi, \label{eq:minima_1}\\
2kx_n^- &=& \pi - \arccos\big[\frac{\cos(\alpha(\tau))}{2}\big] + 2n\pi. \label{eq:minima_2}
\end{eqnarray}

\subsubsection{Harmonic approximation}

We apply a harmonic approximation to the optical potential around each local minimum. Taylor expansion of the lattice potential with respective to the minimum point $x_m$ gives
\begin{eqnarray}
V_{OL} &\approx& V_{OL}(x_m,\tau) + \frac{\partial V_{OL}(x,\tau)}{\partial x}|_{x=x_m} (x-x_m)
+\frac{1}{2!} \frac{\partial^2 V_{OL}(x,\tau)}{\partial x^2}|_{x=x_m} (x-x_m)^2.
\end{eqnarray}
Using the equation for $V_{OL}(x,\tau)$ in Eq.~(\ref{eq:V_OL}) and  $x_m=x_n^+, x_n^-$ from Eqs.~(\ref{eq:minima_1}) and (\ref{eq:minima_2}), respectively, we have
\begin{eqnarray}
V_{OL}(x_m,\tau)  &=& V_0 \sin^2\alpha, \label{eq:V_0th}\\
\frac{\partial V_{OL}(x,\tau)}{\partial x}|_{x=x_m} &=&0, \label{eq:V_1st}\\
\frac{1}{2!} \frac{\partial^2 V_{OL}(x,\tau)}{\partial x^2}|_{x=x_m} &=& 4k^2V_0 (4-\cos^2\alpha), \label{eq:V_2nd}
\end{eqnarray}
allowing the optical potential around a local minimum to be approximated by
\begin{eqnarray}
V_{OL}(x,\tau) &=& V_0 \sin^2\alpha + 4k^2V_0 (4-\cos^2\alpha) (x-x_m)^2.
\end{eqnarray}
Here the dependence of $V_{OL}$ on the index $n$ via $x_m$ implies that all local potentials have identical form. 

Using the general quadratic form of potential energy for harmonic oscillators, 
\begin{eqnarray}
\frac{1}{2} \mu\omega^2(x-x_m)^2 &=& \frac{1}{2} \frac{\partial^2 V_{OL}(x,\tau)}{\partial x^2}|_{x=x_m} (x-x_m)^2 ,
\end{eqnarray}
with $\mu$ the mass of the oscillator, enables us to define a vibrational frequency
\begin{eqnarray}
\omega &=& \sqrt{\frac{\hbar}{\mu} \frac{\partial^2 V_{OL}(x,\tau)}{\partial x^2}|_{x=x_n^+,x_n^-}} 
= \sqrt{\frac{8V_0k^2}{\mu} (4-\cos^2\alpha)} ,
\end{eqnarray}
and zero-vibration amplitude
\begin{eqnarray}
r_0 &=& \sqrt{\frac{1}{\mu\omega}} = [8V_0k^2 \mu (4-\cos^2\alpha)]^{-1/4} .
\label{eq:r_0}
\end{eqnarray}

\subsubsection{Wannier states}

Building on the above harmonic approximation, we employ the ground state wavefunction of the harmonic oscillator, i.e., $u_0(x)=\langle x|0\rangle=(\frac{m\omega}{\pi \hbar})^{1/4} e^{-\frac{m\omega x^2}{2\hbar}}$ 
to construct localized Wannier states for our SSH lattice (setting $\hbar=1$):
\begin{eqnarray}
\chi_0(x) &=& (\frac{\mu\omega}{\pi})^{1/4} e^{-\frac{\mu\omega x^2}{2}}
\overbrace{=}^{Eq.~(\ref{eq:r_0})} \frac{1}{r_0^{1/2} \pi^{1/4}} e^{-x^2 / 2r_0^2}, 
\label{eq:Wannier}
\end{eqnarray}
or equivalently, 
\begin{eqnarray}
\chi_n^+ (x-x_n^+) &=&  \frac{1}{r_0^{1/2} \pi^{1/4}} e^{-(x-x_n^+)^2 / 2r_0^2}, \\
\chi_n^- (x-x_n^-) &=&  \frac{1}{r_0^{1/2} \pi^{1/4}} e^{-(x-x_n^-)^2 / 2r_0^2} .
\end{eqnarray}

\subsubsection{Hopping amplitudes}

Using the Wannier states of Eq.~(\ref{eq:Wannier}), the hopping amplitude can then be calculated explicitly as
\begin{eqnarray}
J_i &=& J_i(\Delta x_i) = \int_{-\infty}^{\infty} dx \chi_0(x+\Delta x_i) \frac{\mu\omega^2 x^2}{2} \chi_0(x) 
= \frac{\mu \omega^2}{2} e^{-\frac{(\Delta x_i)^2}{4 r_0^2}} \big[ \frac{(\Delta x_i)^2}{4} + \frac{r_0^2}{2} \big] .
\end{eqnarray}
Here $\Delta x_1$ and  $\Delta x_2$ are the intracell and intercell distances, respectively, which are obtained from Eqs.~(\ref{eq:minima_1}) and (\ref{eq:minima_2}) as
\begin{eqnarray}
\Delta x_1 (\alpha) &=& x_n^+ - x_n^- = \frac{1}{2k} (2 \arccos\frac{\cos\alpha}{2}), \\
\Delta x_2 (\alpha) &=& x_{n+1}^- - x_n^+ = \frac{1}{2k} (2\pi - 2 \arccos\frac{\cos\alpha}{2}) 
= \Delta x_1 (\pi-\alpha).
\end{eqnarray}
Introducing dimensionless variables $\Delta_i = \frac{\Delta x_i}{2 r_0}$ 
(i.e., the ratio between the intracell (intercell) distances and the zero-vibration amplitude) and using $\mu\omega r_0^2=1$ from Eq.~(\ref{eq:r_0}), allows the hopping amplitudes to be reduced to the simple form
\begin{eqnarray}
J_i &=& \frac{\omega}{2} e^{-\Delta_i^2} (\Delta_i^2 + \frac{1}{2}) ,
\end{eqnarray}
where
\begin{eqnarray}
\Delta_1 (\alpha) &=& \frac{2k\Delta x_1 (\alpha)}{2 k r_0 (\alpha)} 
= \arccos(\frac{\cos\alpha}{2}) \big[\frac{8V_0\mu}{k^2}  (4-\cos^2\alpha) \big]^{1/4} , \\
\Delta_2 (\alpha) &=& \frac{\Delta x_2 (\alpha)}{2 r_0 (\alpha)}
= \frac{2\pi - 2k\Delta x_1 (\alpha)}{2k} \frac{1}{2r_0} = \Delta_1 (\pi-\alpha). 
\end{eqnarray}
This completes the derivation of the hopping amplitudes in Eq.~(2) of the main text.

\begin{figure}
\centering
  \includegraphics[width=.9\columnwidth]{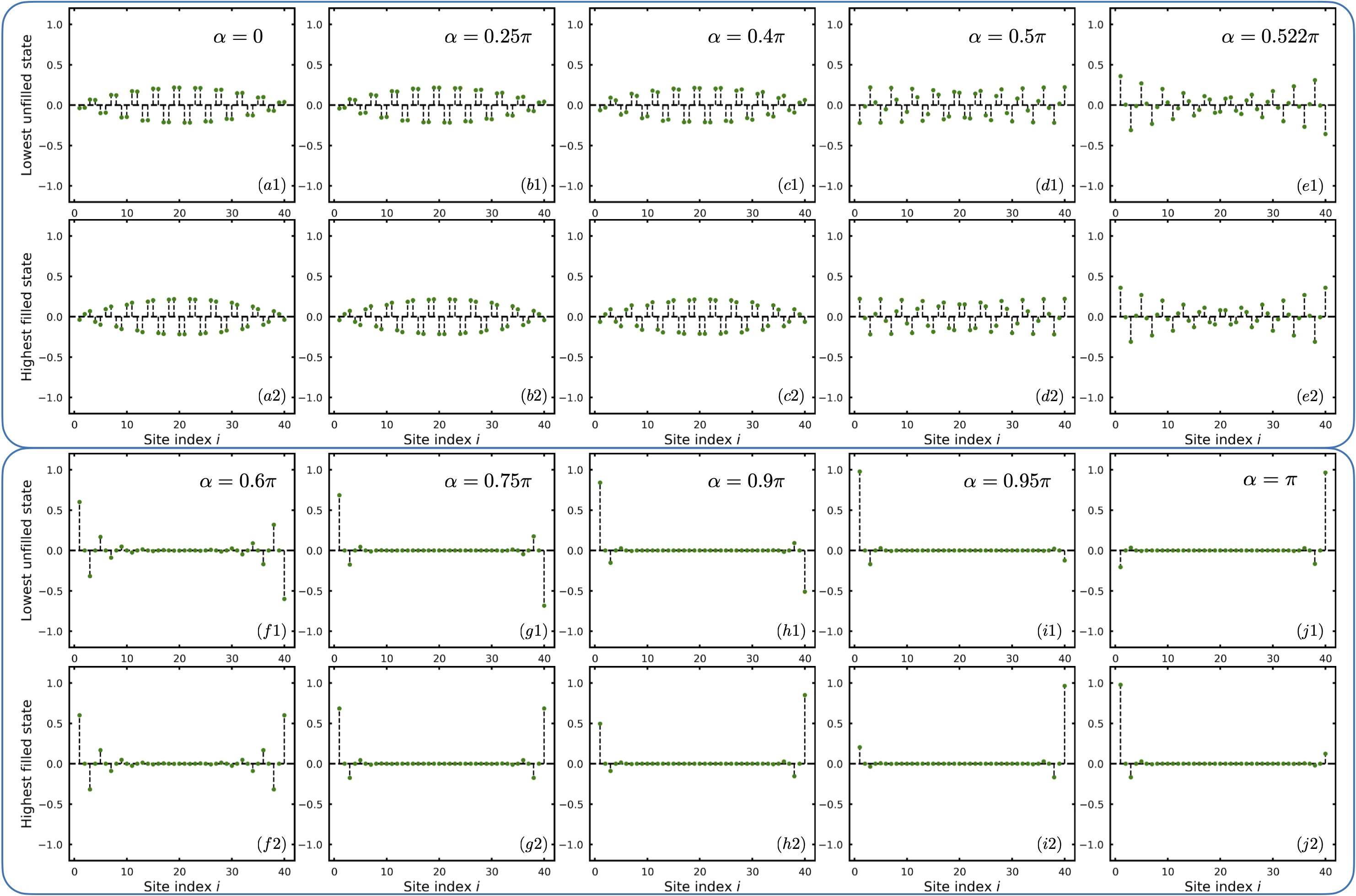}
\caption{(color online) Distribution of the lowest unfilled and highest filled eigenstates for various values of $\alpha$ for the $2N=40$ finite size SSH lattice. All parameters are the same as in Fig.1 of the main text.}
\label{fig:eigenstates}
\end{figure}

\section{Eigenstates of the SSH lattice}

 Fig.~1(b) of the main text shows the energy spectrum as a function of the relative phase parameter $\alpha$ of the three lasers defining the optical lattice. Here we demonstrate several examples of corresponding eigenstates of the finite SSH lattice in trivial, gapless, and topological phase regimes in Fig.~\ref{fig:eigenstates}. All plots here are for $2N =40$ sites.

In the trivial phase the unit cells are weakly coupled to each other and excitations are mostly localized, leading to vanishing disconnected entanglement entropy ($S^D=0$ as shown in Fig.~2(a) of the main text). Fig.~\ref{fig:eigenstates} shows the expected dimerized structure in both the lowest unfilled state and highest filled state for $\alpha = 0, 0.25\pi, 0.4\pi$.  The difference between these two states at a given value of $\alpha$ will be either a positive (Fig.~\ref{fig:eigenstates} (a1), (b1), or (c1)) or negative (Fig.~\ref{fig:eigenstates} (a2), (b2), or (c2)) superposition of the two site states in a unit cell. 

With regard to the gapless phase of the lattice, we consider here $\alpha=0.5\pi$ as well as several points around this on either side, since for a finite chain this phase is not sharply defined at a single $\alpha$ value. 
It is evident that the lowest unfilled state at  $\alpha = 0.5\pi$ (Fig.~\ref{fig:eigenstates} (d1), with $S^D/\log2 \sim 0.177$) is similar to that of $\alpha=0.4\pi$, for which $S^D=0$ (Fig.~\ref{fig:eigenstates} (c1)).  However this state at $\alpha = 0.5\pi$ is quite different from that the corresponding lowest unfilled state at $\alpha= 0.6 \pi$, for which $S^D/\log2 =2$.
In contrast, the highest filled state at $\alpha = 0.5\pi$ (Fig.~\ref{fig:eigenstates} (d2)) is quite different from that at $\alpha=0.4\pi$,  for which $S^D=0$ (Fig.~~\ref{fig:eigenstates} (c2)), but is very similar to that  of $\alpha = 0.522\pi$ for which $S^D/\log2 \sim 1$  (Fig.~\ref{fig:eigenstates} (e2)), as well to that of $\alpha = 0.6 \pi$  for which $S^D/\log2 =2$. 
In other words, the lowest unfilled state at $\alpha = 0.5\pi$ appears to be obtainable by smoothly increasing $\alpha$ from a value in the trivial phase, while the highest filled state at $\alpha = 0.5\pi$ appears smoothly connected to the corresponding state in the topological phase on varying $\alpha$. In particular, the similarity between Figs.~\ref{fig:eigenstates} (d2) and (e2) enables the point $\alpha = 0.5\pi$ to be referred as a gapless phase for the $2N=40$ finite lattice.

In the topological phase ($S^D/\log2=2$), the lowest unfilled and highest filled states become degenerate (see Fig.~1(b) of the main text). At $\alpha = 0.6\pi$ or $0.75\pi$, these states resemble Bell states, i.e.,  anti-symmetric (Figs.~\ref{fig:eigenstates} (f1) or (g1)) and symmetric (Figs.~\ref{fig:eigenstates} (f2) and (g2)) superposition of edge states with equal weight. When further increasing $\alpha$ (e.g., to $0.9\pi$, or $0.95\pi$), each eigenstate approaches a single edge-site state (see Fig.~\ref{fig:eigenstates} (h1), (h2) or (i1), (i2)). In particular, panels (j1) and (j2) for $\alpha = \pi$ provide good approximations to localized edge-site states.

\begin{figure}
\centering
  \includegraphics[width=.48\columnwidth]{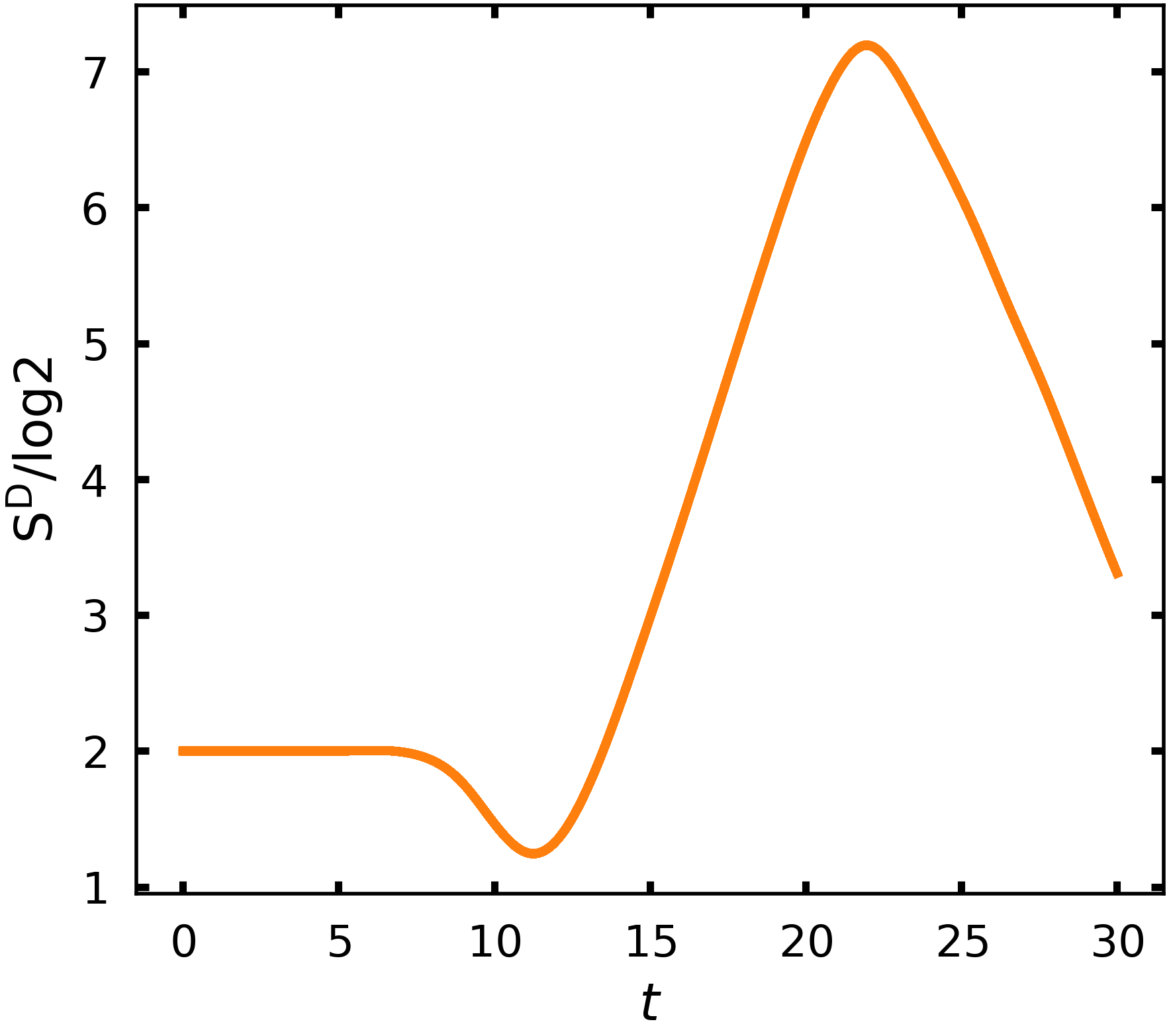}
\caption{(color online) Dynamics of disconnected entanglement entropy $S^D(t)$ for an SSH lattice of $2N=40$ sites. The evaluation of $S^D(t)$ is performed in a similar way as in Fig.2(a) of the main text, i.e., by using an even partition of the $2N$ sites into four non-overlapping segments.}
\label{fig:EE_dynamics}
\end{figure}

\section{Dynamics of entanglement entropy}

In the main text the time independent entanglement entropy $S^D$ is employed to characterize the finite size analog of the bulk topological phase transition (see Fig.~2 of the main text).  
Here we extend this analysis to consider the time-dependence of the entanglement entropy under quantum quench. 

We first give a brief introduction to the calculation in the time-independent setting, following the procedures outlined in Ref.~\onlinecite{Micallo20spp}. 
The basic idea is to use a correlation matrix to compute the reduced density matrix $\rho_{X}$ of a subsystem $X$ via the relation of this with an entanglement Hamiltonian ${\mathcal H}_X$. 
The correlation matrix can be defined as
\begin{eqnarray}
(C_X)_{mn}(0) &=& \sum_{k<k_{\rm F}} \Psi_k^*(m)\Psi_k(n) ,
\label{eq:C_X}
\end{eqnarray}
where $\Psi_k$  are the single-particle  eigenstates of the lattice Hamiltonian with site indices $m,n$ of the subsystem $X$. 
Making use of the relation of $C_X$ to the entanglement Hamiltonian ${\mathcal H}_X$~\cite{Peschel03jpa} 
\begin{eqnarray}
{\cal H}_X &=& {\rm log}\frac{1-C_X}{C_X} ,
\end{eqnarray}
we then readily obtain the reduced density matrix 
\begin{eqnarray}
\rho_X &=& \frac{e^{-{\cal H}_X}}{{\rm Tr}_X [e^{-{\cal H}_X}]} ,
\end{eqnarray}
and hence the bipartite entanglement entropy 
\begin{eqnarray}
S_X &=& -{\rm Tr}_X[\rho_X [\log(\rho_X)] .
\end{eqnarray}
Computationally, we construct Eq.~(\ref{eq:C_X}) and numerically diagonalize this to obtain the spectrum of $C_X$ from which we then compute $S^D$. 
For the case considered in the main text, we evaluate $S^{D}=S_{AB}+S_{BC}-S_{ABC}-S_{B}$ with $A,B,C,D$ being four non-overlapping segments of the SSH lattice, each of which has $N/2$ sites. 

To study the entanglement entropy dynamics, we consider a similar scenario to that in Ref.~\onlinecite{Micallo20spp}. 
Specifically, we focus on the case of two excitations without dissipation ($\gamma=0$) or pumping ($F_A=F_B=0$), ensuring conservation of the number of particles, and further consider the quench to be implemented at $t=0$.
After preparation in the ground state of $H_{\rm SSH}(\alpha_{\rm T})$ with $\alpha_{\rm T}=0.75\pi$ for $t<0$, a quantum quench is applied at $t=0$, and the SSH lattice then evolves under the new Hamiltonian $H_{\rm SSH}(\alpha_{\rm G})$ with $\alpha_{\rm G}=0.5\pi$ for $t>0$.
The time-dependent correlation matrix $C_X(t)$ for the state $\rho(t)$ can be written in terms of the eigenvalues $E_K$ and eigenvectors $\Phi_k$ of $H_{\rm SSH}(\alpha_{\rm G})$,
\begin{eqnarray}
(C_{X})_{mn}(t) &=& {\rm Tr}[\rho(t)c_m^{\dagger}(0)c_n(0)]
={\rm Tr}[\rho(0)c_m^{\dagger}(t)c_n(t)] \notag\\
&=&
\sum_{k,k',m',n'} \Phi_k^*(m) \Phi_{k'}(n) e^{-iE_{k'}t}e^{iE_kt}
(C_X)_{m'n'}(0) \Phi_{k'}^*(n')\Phi_k(m'),
\end{eqnarray}
with $c_m^{\dagger}(t)$ ($c_m(t)$) being the Heisenberg representation of $c_m^{\dagger}$ ($c_m$).
Thus $(C_X)_{mn}(0)=\sum_{k<k_{\rm F}} \Psi_k^*(m)\Psi_k(n)$, as given in Eq.~(\ref{eq:C_X}), where $\Psi_k$ are ground states of $H_{\rm SSH}(\alpha_{\rm T})$.
We can then evaluate $S^D(t)$ from the spectrum of $C_X(t)$. 
Fig.~\ref{fig:EE_dynamics} shows a plot of the resulting time dependence of the entanglement entropy, $S^D(t)$. It is evident that for a short time scale $S^D$ remains at or close to its initial value $S^D = 2\log2$ in the topological phase ($\alpha = 0.75\pi$)~\cite{Micallo20spp}.

\begin{figure}
\centering
  \includegraphics[width=.85\columnwidth]{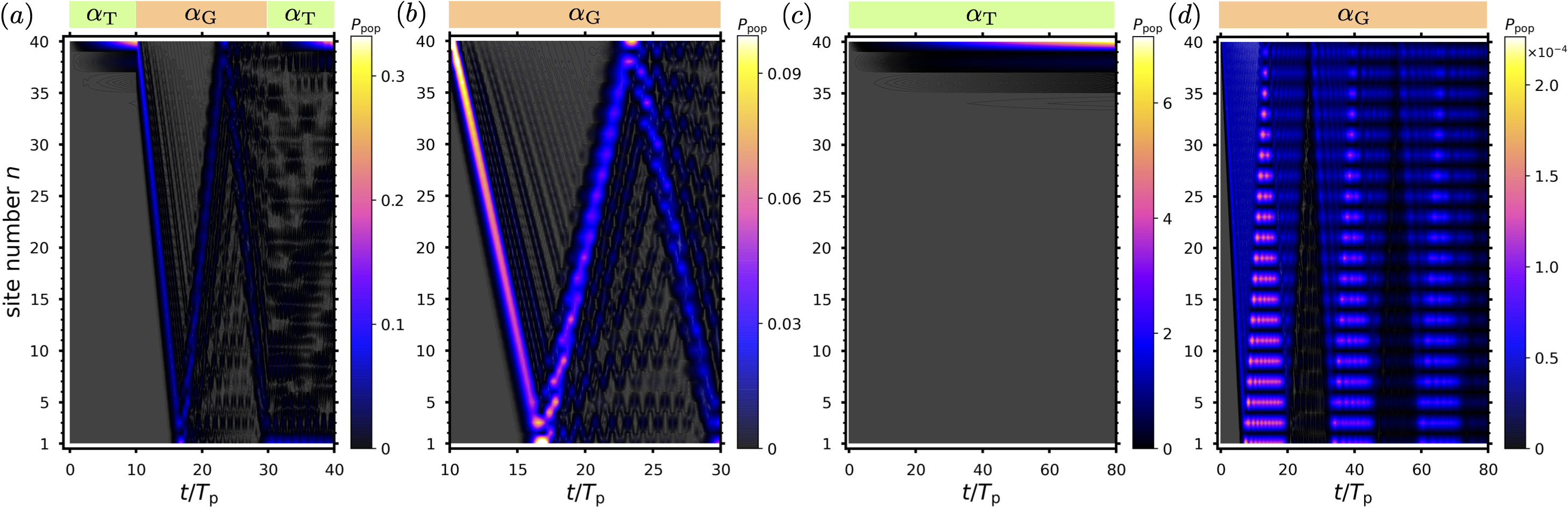} 
\caption{Time-dependent occupation number $P_{\rm pop}$ for the case of many excitations with a pumping field applied at the last site  ($F_{\rm A}=0$, $F_{\rm B}=0.01$) of the finite SSH lattice with $2N=40$ sites. Parameters are the same as in Fig. 3 of the main text.}
\label{fig:pumpN_initialOccupNo_damp}
\end{figure}

\begin{figure}
\centering
  \includegraphics[width=.85\columnwidth]{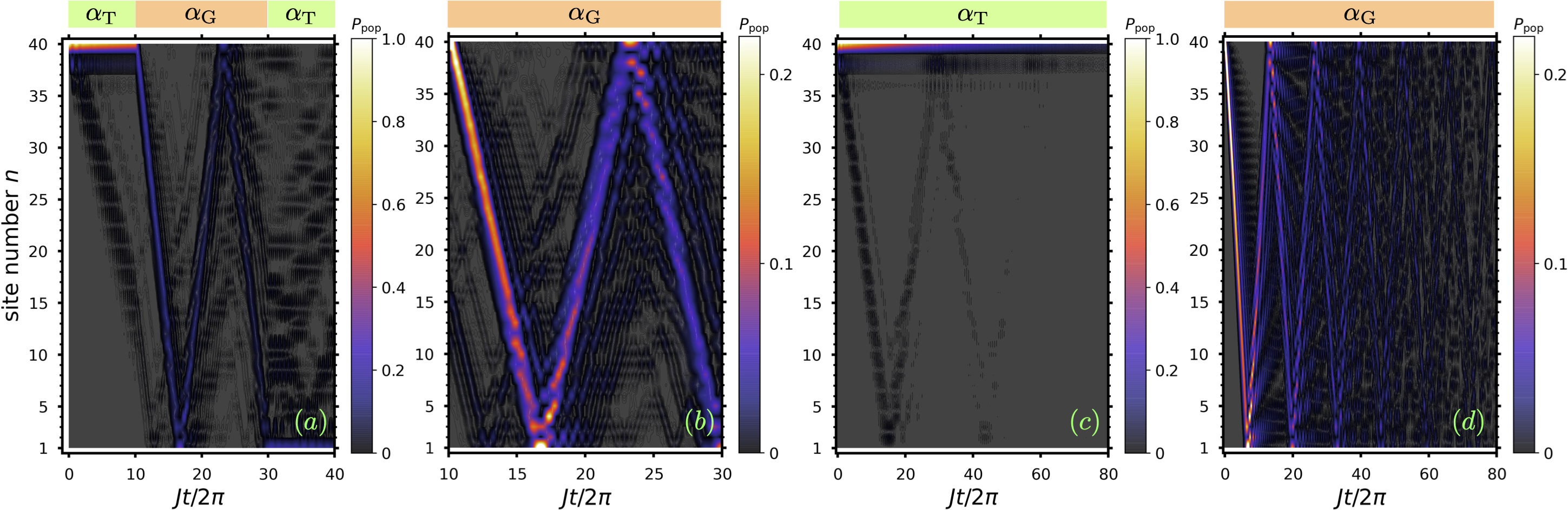} 
\caption{Time-dependent occupation number $P_{\rm pop}$ for the case of an excitation initially occupying the last site ($A_{1}(0)=0$, $B_{20}(0)=1$ ) and no pumping.
Parameters are the same as in Fig.~9 of the main text.}
\label{fig:pumpN_initialOccupYes_damp}
\end{figure}

\section{Time evolution under pumping of the last site of the finite SSH chain}

In the main text we have shown results for the  pumping fields applied at both end sites and also for a pumping field applied only at the first site. For completeness, we present here results for pumping at the last site.  Figs.~\ref{fig:pumpN_initialOccupNo_damp} and \ref{fig:pumpN_initialOccupYes_damp} show the time dependence of the site occupation numbers for this pumping in the case of many and two excitations, respectively. 
The figures show that there is a reflection symmetry with respect to the lattice center between the results shown here 
for pumping of the last site, and the corresponding results pumping of the first site in 
Figs. 3, 6, 9 and 12 of the main text.




\begin{figure}
\centering
  \includegraphics[width=.85\columnwidth]{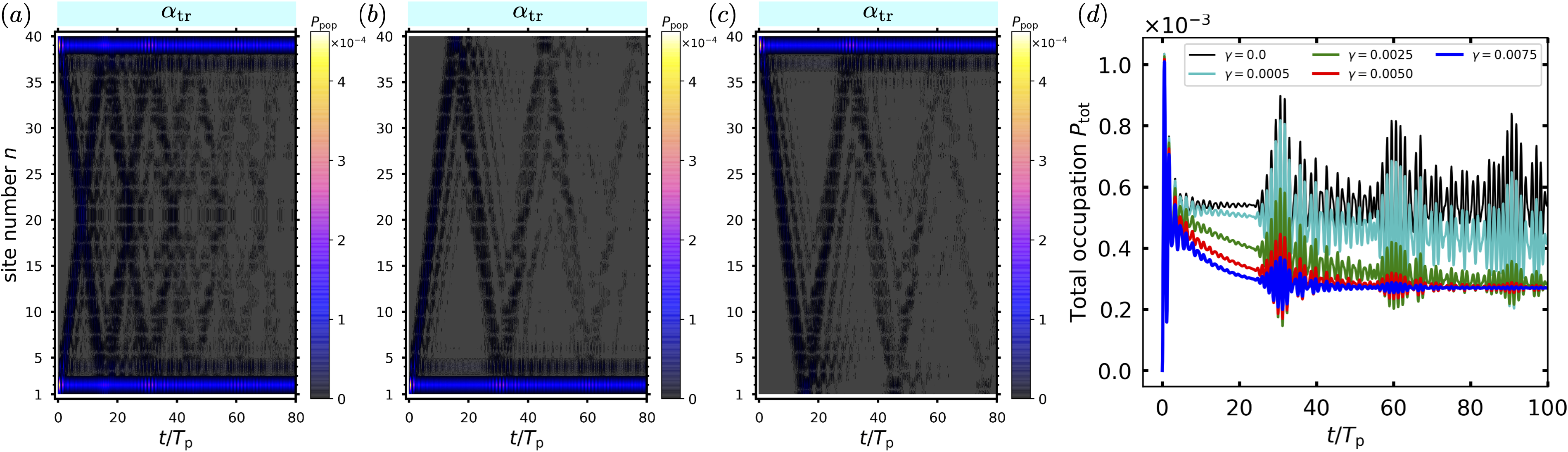} 
\caption{Time evolution of the occupation number $P_{\rm pop}$ for many excitations pumped from (a) both ends, (b) the first site, and (c) the last site of the SSH lattice in the trivial phase regime with $\alpha_{\rm tr}=0.25\pi$. In this regime, large populations appear on the pumping sites, i.e., the $1$st and/or last sites and there is still a small number of excitations that hop between adjacent sites. Here we take $\gamma=0.0025$. (d) Time dependence of the total number excitations $P_{\rm tot}$ in the case of pumping from both ends (panel (a)) in the trivial phase regime. Other parameters are same as in Fig.~6 of the main text.}
\label{fig:trivial_initialOccupNo_damp}
\end{figure}

\begin{figure}
\centering
  \includegraphics[width=.8\columnwidth]{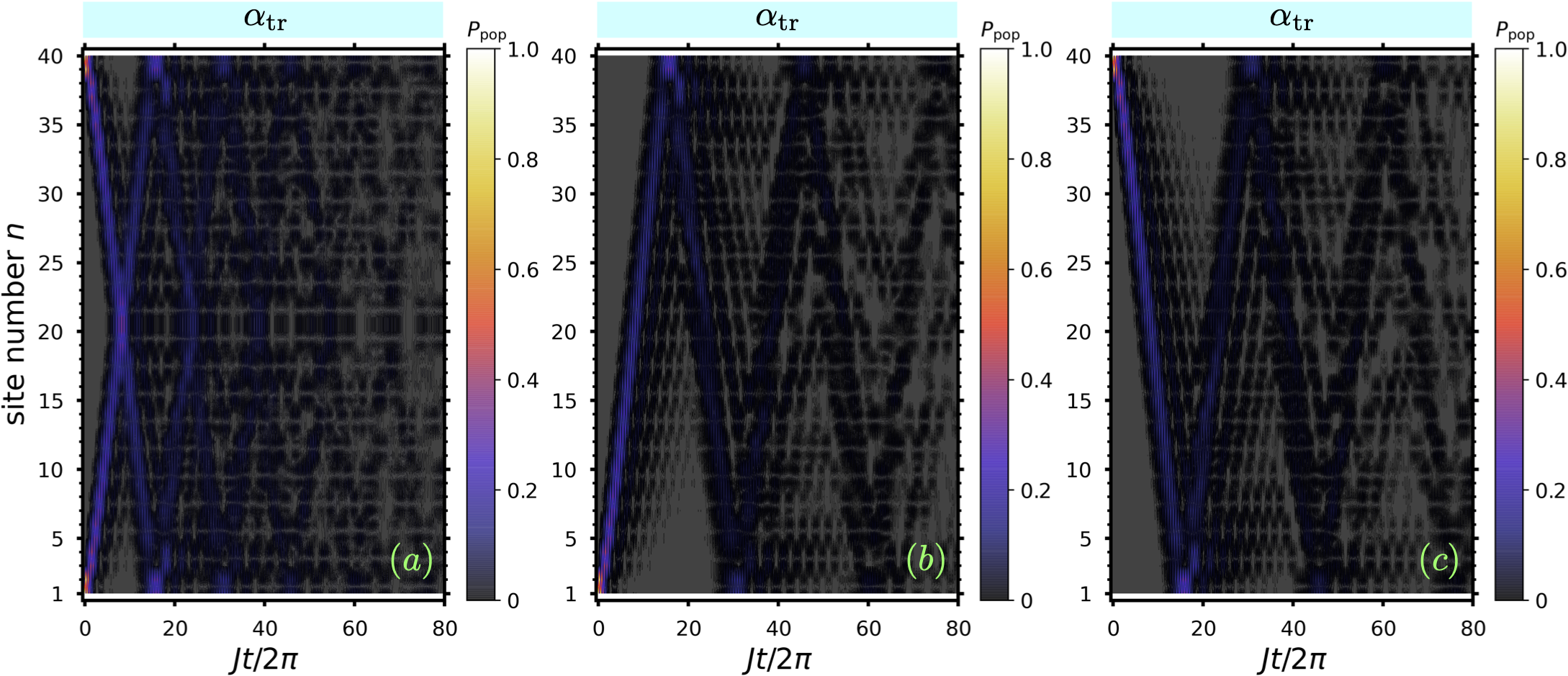} 
\caption{Time evolution of the occupation number $P_{\rm pop}$ for (a) two excitations initially at both ends and one excitation initially occupying (b) the first site and (c) the last site of the SSH lattice in the trivial phase regime with $\alpha_{\rm tr}=0.25\pi$. An interference pattern due to non-topological states appears in (a). Here we set the damping parameter to $\gamma=0.0025$. All other parameters are the same as in Fig.~12 of the main text.}
\label{fig:trivial_initialOccupYes_damp}
\end{figure}

\section{The trivial phase regime without quench}

To complement the analysis of topological and gapless phase regimes without quench that were considered in the main text in order to distinguish features of topological interference, here we add the presentation of results in the trivial phase regime wihout quench. The site occupation number patterns in this regime, shown in Figs.~\ref{fig:trivial_initialOccupNo_damp} and \ref{fig:trivial_initialOccupYes_damp} for the many and two/one excitations respectively, are quite different from both the topological interference patterns and the individual patterns in the topological and gapless phases that were presented in the main text. 
We additionally note that, as was observed for the gapless phase in the main text, comparison between Fig.~\ref{fig:trivial_initialOccupNo_damp} and Fig.~\ref{fig:trivial_initialOccupYes_damp} shows a dependence of the site occupation number pattern in the trivial phase on the initial number of the excitations in the lattice.

\begin{figure}
\centering
  \includegraphics[width=.4\columnwidth]{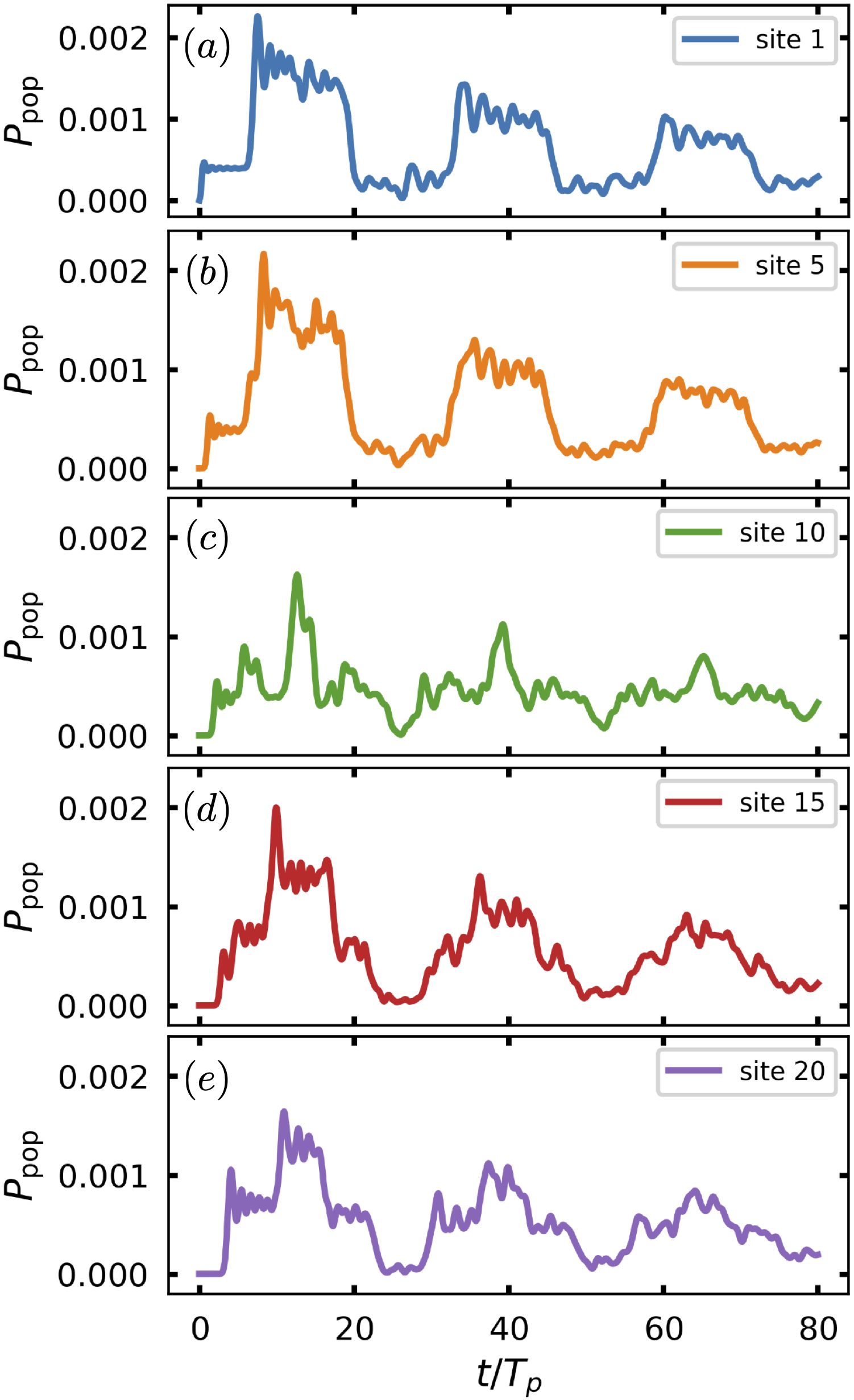}
\caption{
The occupation number $P_{\rm pop}$ at five specific sites (a) $1$, (b) $5$, (c) $10$, (d) $15$, and (e) $20$ of Fig.~6(b) of the main text in the gapless phase. The parameters are same as in Fig.~6(b). }
\label{fig:popu_sites_5_10_15_20_gapless}
\end{figure}

\section{Population evolution at specific sites in the gapless phase}

Fig.~\ref{fig:popu_sites_5_10_15_20_gapless} shows the time dependent of the site occupation probability at specific sites in Fig.~6(b) of the main text. The results show oscillatory behavior at all sites and as well as nonzero population at both odd and even sites, which explains the observations in Fig.~6(b) as well as Fig.~7(c) and (d) of the main text and supports the discussion in the main text.

\begin{figure*}
\centering
  \includegraphics[width=0.4\columnwidth]{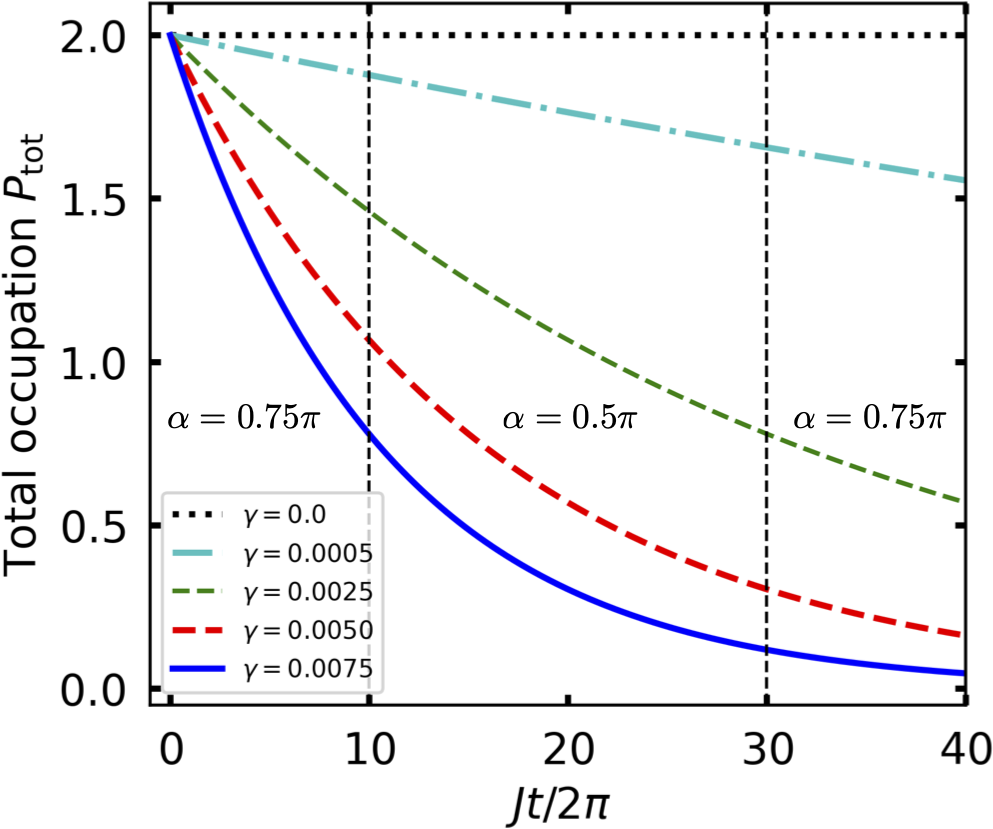} 
\caption{
Evolution of the total number $P_{\rm tot}$ of excitations for the two-excitation case with various values of damping rate $\gamma$ under the quench scheme of the main text that generates topological interference in the region $10 < Jt/2\pi < 30$. 
The curve with $\gamma=0.0025$ corresponds to Fig.~9(a) under the quantum quench protocol in the main text. Other parameters are the same as Fig.~9 of the main text.}
\label{fig:twopaticle_totalOccup_damp_interfer}
\end{figure*}

\begin{figure}
\centering
  \includegraphics[width=0.4\columnwidth]{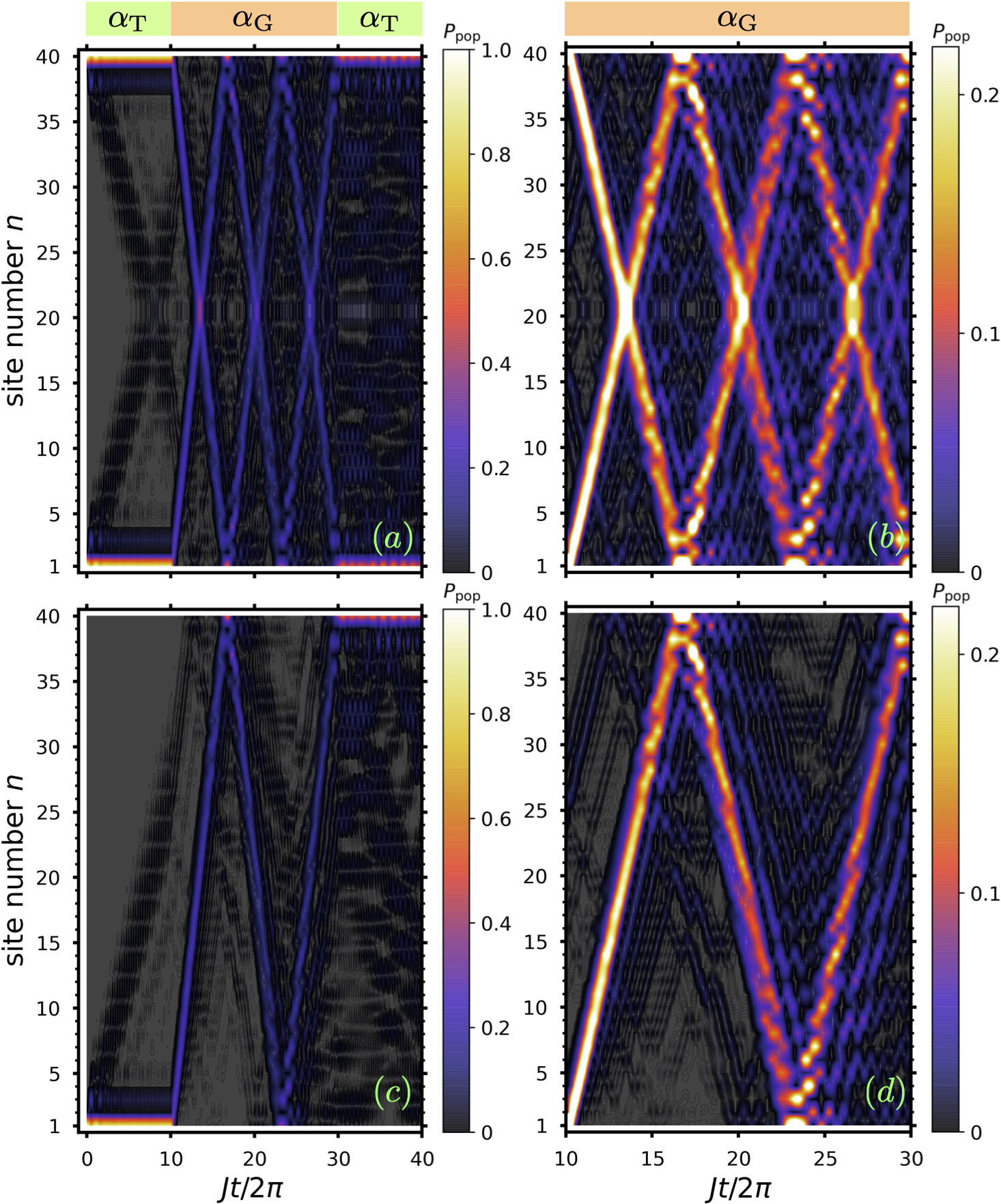}
\caption{
Time-dependent occupation number $P_{\rm pop}$ in the case of two excitations (top panels) and of one excitation (bottom panels) inserted in the SSH lattice at $t=0$.
Panels (a) - (b): the two excitations initially occupy both ends ($A_{1}(0)=B_{20}(0)=1$). Panels (c) - (d): one excitation initially occupies the first site ($A_{1}(0)=1$, $B_{20}(0)=0$).
Topological interference is observed in (a) with $\alpha_{\rm T}=0.75\pi$, $\alpha_{\rm G}=0.5\pi$, and the switching time $Jt_a/2\pi=10T_p$, $Jt_b/2\pi=30$.
Panels (b) and (d) are zoom-in views of the patterns during $10<Jt/2\pi<30$ in (a) and (c), respectively.
Here the pumping fields are switched off ($F_A=F_B=0$).
We take $\gamma=0$ and other parameters are same as in Fig.~9 of the main text. 
}
\label{fig:twoparticle_gamma_0}
\end{figure}

\begin{figure}
\centering
  \includegraphics[width=0.4\columnwidth]{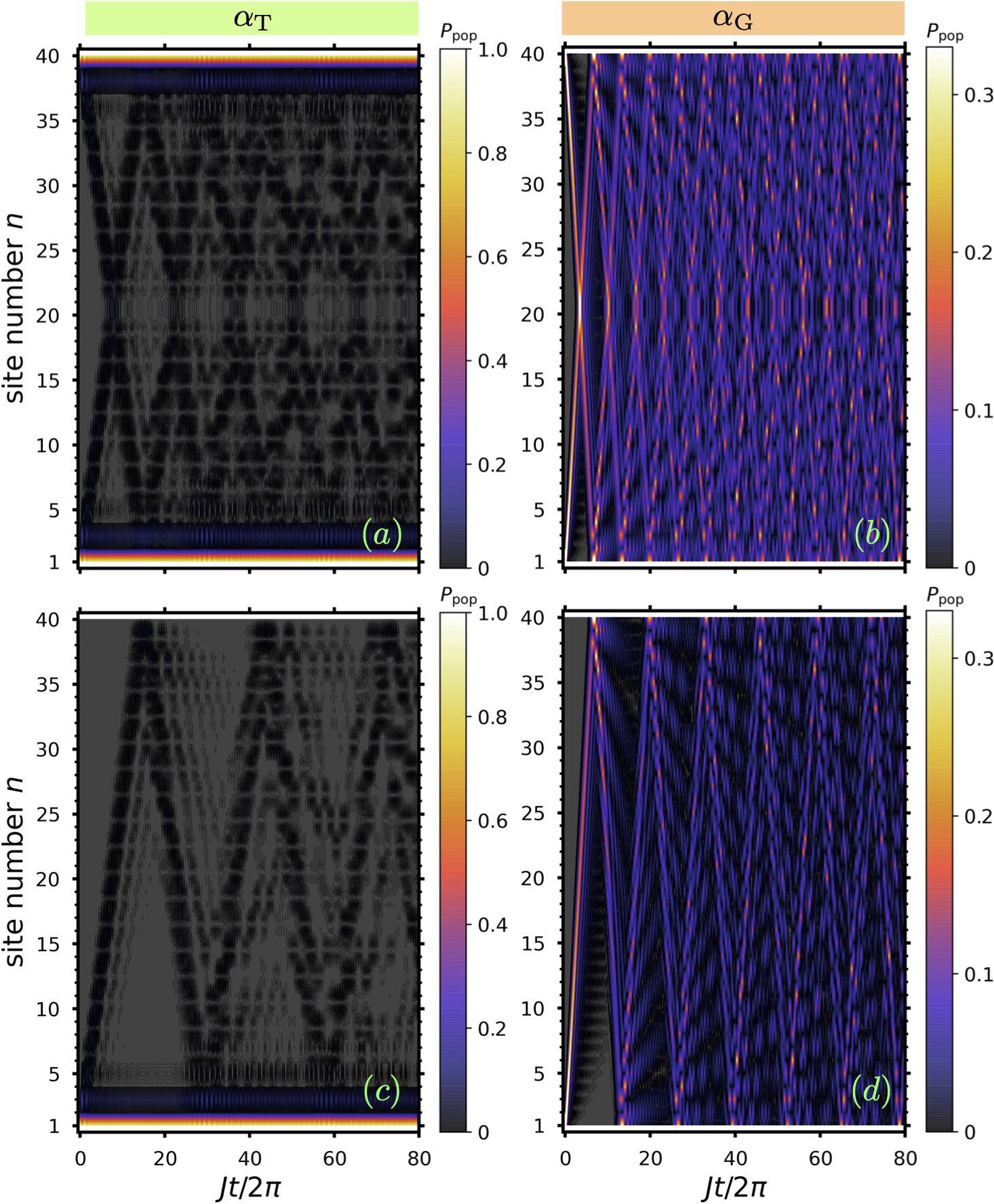}
\caption{
Time-dependent occupation number $P_{\rm pop}$ in the case of two excitations (top panel) and one excitation (bottom panels) inserted at $t=0$, for the topologically nontrivial phase (left panels, $\alpha_{\rm T}=0.75\pi$) and for the gapless phase (right panels, $\alpha_{\rm G}=0.5\pi$).
Panels (a)-(b): the two excitations initially occupy both ends ($A_{1}(0)=B_{20}(0)=1$). Panels (c)-(d): one excitation occupies the first site ($A_{1}(0)=1$, $B_{20}(0)=0$).
Here the pumping fields are switched off ($F_A=F_B=0$). We take $\gamma=0$ and all other parameters are same as in Fig.~12 of the main text. 
}
\label{fig:twoparticle_gamma_0_B}
\end{figure}

\section{Damping effects}

To support our prediction that the topology-independent damping in each phase (see Fig.~13(d) in the main text) is valid also 
in the case of the presence of the phase switching, i.e., of the quantum quench
we show in Fig.~\ref{fig:twopaticle_totalOccup_damp_interfer} the time evolution of the total population of the lattice in the case of two excitations at $t=0$, under several values of the damping rate $\gamma$). It is evident that Fig.~\ref{fig:twopaticle_totalOccup_damp_interfer} shows a smooth decay as the quench proceeds, similar to that seen in Fig.~13(d) of the main text for calculations without phase switching, implying that the universal topology-independent damping still holds in the quantum quench situation 

However, unlike the topology-independence, i.e., $\alpha$-independence, of the decay of the total population of excitations under damping presented above and in the main text, the {\it interference} of the topological states under quenching could be suppressed by the decay differently than the interference patterns of unquenched states. To illustrate this, we show in Figs.~\ref{fig:twoparticle_gamma_0} and \ref{fig:twoparticle_gamma_0_B}  calculations with no damping, i.e., $\gamma=0$, that are otherwise analogous to Figs.~9 and 12 of the main text ($\gamma = 0.0025$).
The greater extent of suppression of interferences of the topological states for the $\gamma=0.0025$ case is evident from the lower range of the color scale in Fig. 9 of the main text, as well as in the decay of the interferences at longer times in the upper right panel of that Figure.



\begin{figure}
\centering
  \includegraphics[width=.4\columnwidth]{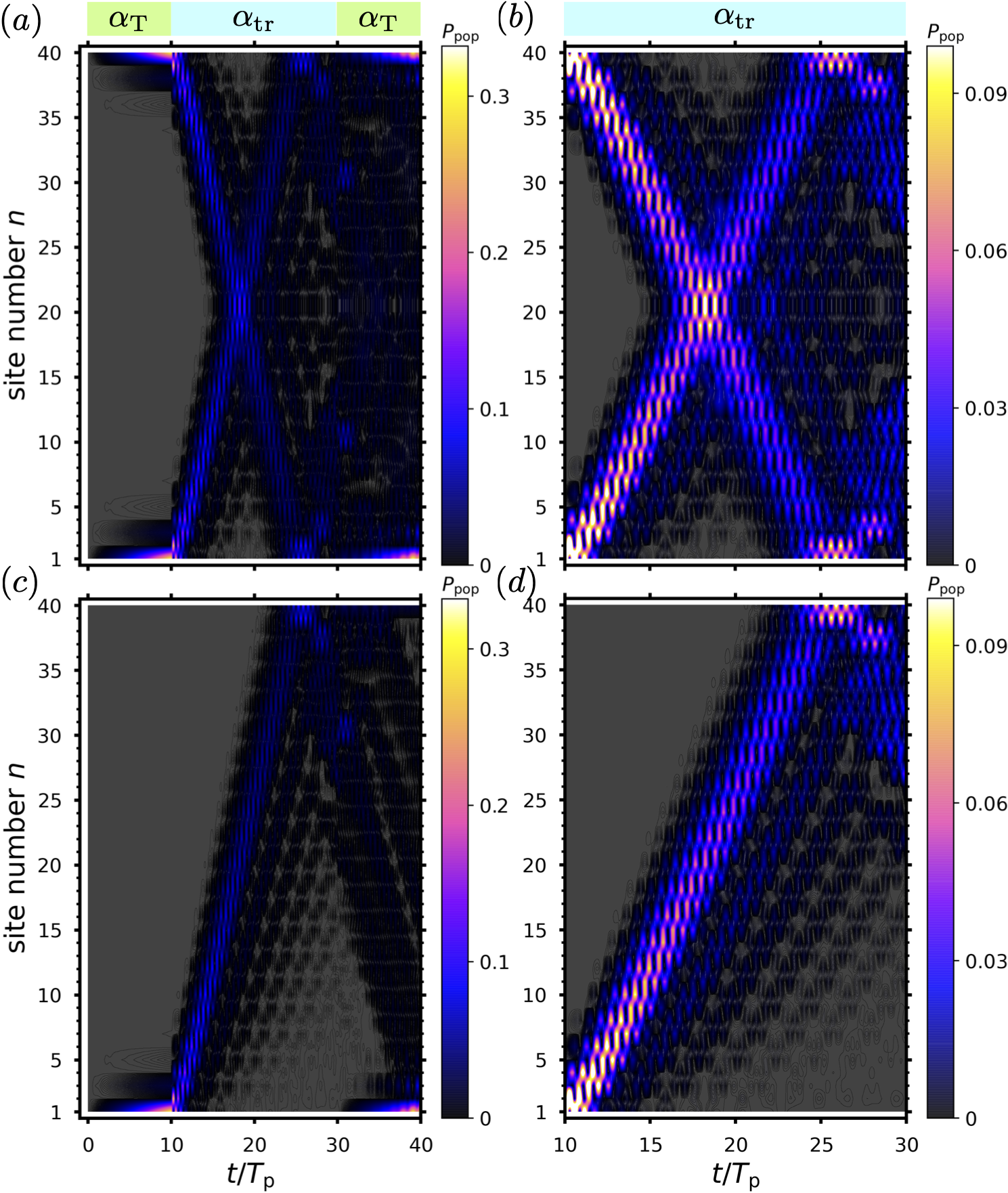}
\caption{Time-dependent occupation number $P_{\rm pop}$ of the SSH lattice during an alternative interference-inducing quantum quench protocol into the trivial phase $\alpha_{\rm tr}$.  The upper panels show results under pumping at both ends [$F_{\rm A}=F_{\rm B}=0.01$], and the lower panels results under pumping at one end only [$F_{\rm A}=0.01$, $F_{\rm B}=0$]. 
The initial lattice condition at $t=0$ is no excitation, i.e., $A_l(0)=B_l(0)=0$.
Topological interference is observed in panels (a) and (c) with $\alpha_{\rm T}=0.75\pi$, $\alpha_{\rm tr}=0.25\pi$, and switching time interval specified by $t_a=10T_p$, $t_b=30T_p$.
Panels (b) and (d) are zoom-in views of the pattern in the trivial phase quench period $10T_p<t<30T_p$ of panels (a) and (c), respectively.
The other parameters used are $\gamma=0.0025$, $N=20$, $\varepsilon=1$, $V_0=0.125$, $\mu/k^2=0.25$, $\omega_{\rm A}=\omega_{\rm B}=1$, $\phi_{\rm A}=\phi_{\rm B}=\phi=0$.}
\label{fig:multiparticle_quenchedTrivial}
\end{figure}

\section{An alternative quantum quench protocol}

In the main text, we have studied the quantum quench protocol that considers a quench from the topological phase to the gapless phase (see. e.g., Fig.~3 in main text). 
One could alternatively consider quenching the SSH lattice from the topological phase to the trivial phase, by switching the $\alpha$ value to $\alpha_{\rm tr}=0.25\pi$. 
The  result of this alternate quench is presented in Fig.~\ref{fig:multiparticle_quenchedTrivial}, with similar graphical representation as in Fig.~3 of the main text.
As expected, we observe localized edge states in the topological phase ($\alpha_{\rm T}=0.75\pi$ for $t/T_{\rm p}\leq10$) and an interference pattern when quenching into the trivial phase after switching to $\alpha_{\rm tr}=0.25\pi$ at $t/T_{\rm p}=10$. 
Comparing with Fig.~3(a) of the main text, we see that under this alternate quench protocol there is a different interference pattern in the period $10<t/T_{\rm p}\leq30$. Furthermore, this interference pattern also differs from the case of the trivial phase without quench (see Fig.~\ref{fig:trivial_initialOccupNo_damp}). 